\documentclass[astrosymb, tighten, fleqn]{cas-dc}

\usepackage{amsmath}
\usepackage{amssymb}
\usepackage{mathtools}
\usepackage{url}
\usepackage{gensymb}

\usepackage{cancel}

\newcommand{\de}{{\rm d}}

\usepackage{siunitx}
\usepackage[authoryear]{natbib}
\usepackage{lineno}

\let\oldequation\equation
\let\oldendequation\endequation

\renewenvironment{equation}
  {\linenomathNonumbers\oldequation}
  {\oldendequation\endlinenomath}

\shorttitle{SEDs of Forming Giant Planets}
\shortauthors{Taylor \& Adams}

\begin{document}


\title[mode = title]{Radiative Signatures of Circumplanetary Disks and Envelopes During the Late Stages of Giant Planet Formation}

\author[1]{Aster G. Taylor}[orcid=0000-0002-0140-4475]
\cormark[1]
\fnmark[1]
\ead{agtaylor@umich.edu}
\affiliation[1]{organization={Department of Astronomy, University of Michigan},
    city={Ann Arbor},
    state={MI},
    postcode={48109}, 
    country={USA}}

\author[2,1]{Fred C. Adams}[orcid=0000-0002-8167-1767]
\affiliation[2]{organization={Department of Physics, University of Michigan},
    city={Ann Arbor},
    state={MI},
    postcode={48109}, 
    country={USA}}

\cortext[cor1]{Corresponding author}
\fntext[fn1]{Fannie and John Hertz Foundation Fellow}

\begin{abstract}
During the late stages of giant planet formation, protoplanets are surrounded by a circumplanetary disk and an infalling envelope of gas and dust. For systems with sufficient cooling, material entering the sphere of influence of the planet falls inward and approaches ballistic conditions. Due to conservation of angular momentum, most of the incoming material falls onto the disk rather than directly onto the planet. This paper determines the spectral energy distributions of forming planets in this stage of evolution. Generalizing previous work, we consider a range of possible geometries for the boundary conditions of the infall and determine the two-dimensional structure of the envelope, as well as the surface density of the disk. After specifying the luminosity sources for the planet and disk, we calculate the corresponding radiative signatures for the system, including the emergent spectral energy distributions and emission maps. These results show how the observational appearance of forming planets depend on the input parameters, including the instantaneous mass, mass accretion rate, semimajor axis of the orbit, and the planetary magnetic field strength (which sets the inner boundary condition for the disk). We also consider different choices for the form of the opacity law and attenuation due to the background circumstellar disk. Although observing forming planets will be challenging, these results show how the observational signatures depend on the underlying properties of the planet/disk/envelope system. 
\end{abstract}

\maketitle

\section{Introduction}\label{sec:intro}

Giant planets are thought to form through the core accretion paradigm (e.g., \citealt{Pollack1996}). During the final stage of this process, when most of the mass is accreted, a majority of the material does not fall directly onto the planet itself, but collects into an accompanying circumplanetary disk. The formation, structure, and evolution of these disks represents a crucial part of the giant planet formation process, as most of the material making up the final planetary mass is processed through the disk. The disk also emits a significant fraction of the luminosity of the forming planet, where this radiation is processed further by the surrounding envelope of incoming material. Forming planets are potentially observable in this later stage of evolution. In order to understand future observations, we must determine the emergent spectral energy distributions (SEDs) for forming planets. Toward this end, we determine the radiative signature of forming planets and calculate how the SEDs depend on the properties of the planet, disk, and envelope. These results will enable the determination of system properties in future observations. 

Within the core accretion paradigm, the planet formation process can be divided into three stages \citep{Pollack1996,Helled2014}. In the first stage, solid material from the circumstellar disk collects into a high-metallicity core with mass $M\simeq\qty{10}{M_\oplus}$. After that threshold is reached, the protoplanet begins to accrete gaseous material from the disk and enters the second stage. Here, the protoplanet gathers a hydrostatically-supported gaseous envelope extending out to the Hill radius. The envelope must cool and contract in order for more material to enter the Hill sphere. This process is slow and provides an important bottleneck in the formation process. After the protoplanetary envelope becomes sufficiently massive, it collapses and the planet can experience accretion at a rapid rate. Jovian planets gain most of their mass during this final stage. Although the mechanisms that lead to the end of this final phase remain uncertain, the background circumstellar disks lose their mass supply on time scales of $\sim\qty{3}{\mega yr}$ \citep{Hernandez2007}, and rapid accretion eventually ceases. This condition, along with the existence of Jovian planets, sets constraints on the possible mass accretion rate. During the third stage of rapid accretion, forming planets reach their brightest luminosities and are most easily observed. The structure of the infalling envelopes and the circumplanetary disks largely determine the spectral appearance of these forming objects.  

Recently, observations of the region surrounding gas giant protoplanets have become feasible. Currently, angular resolutions have begun to approach the size scale of circumplanetary disks, enabling direct detections. Although only a few tentative detections have been made \citep{Isella2019, Benisty2021, Bae2022, Christiaens2024, Cugno2024}, more are expected in the coming years. Corresponding models are therefore necessary to connect our understanding of the structure of circumplanetary disks to future observations. 

A great deal of previous work on the late stages of giant planet formation has been carried out. Numerical investigations of the final stage of accretion (e.g., \citealt{Tanigawa2002, Szulagyi2016, Zhu2016,  Szulagyi2017, Lambrechts2019, Schulik2019, Schulik2020, Maeda2022}) show that disks will readily form around protoplanets as a result of angular momentum conservation, provided that the incoming gas can cool sufficiently (see \citealt{Machida2008, Szulagyi2017, Fung2019}). These simulations provide working estimates for the structure of the disks and surrounding envelopes. Nonetheless, current numerical simulations are often limited in temporal and spatial resolution and are not fully consistent with one another. One particular source of disagreement concerns the geometric distribution of the material entering the Hill sphere. Some studies show infall through the midlatitudes \citep{Li2023}, some show primarily equatorial infall (e.g., \citealt{Ayliffe2009}), and others show that material primarily enters through the rotational poles of the system (e.g., \citealt{Lambrechts2017}). 

Previous work \citep{Adams2022,Taylor2024} constructed a semi-analytical treatment for the structure of protoplanets during their final stage of formation. These previous results determine the velocity field and density structure of the infalling envelope, the surface density of the circumplanetary disk, a treatment of the inner boundary condition due to magnetic field truncation, and the luminosity sources for the system. However, the resulting SEDs of the forming planets  were determined under the assumptions that (i) the envelope is spherically symmetric, (ii) the disk is a point source at the center of the envelope, and (iii) the envelope is optically thin to its own radiation.  

{In this paper}, we {extend the aforementioned work by} generalizing the treatment of radiative transfer to include the full two-dimensional structure of the envelope and {the finite extent of the} circumplanetary disk (see also \citealt{Choksi2024, Sun2024}). We determine the temperature structure of the envelope using two complementary treatments. First, we use a simple parametric form for the temperature profile and employ global conservation of luminosity to set the parameters. Second, we use a Monte-Carlo radiative transfer code \citep{Dullemond2012} that self-consistently determines the temperature. The two approaches are in excellent agreement, provided that the optical depth of the envelope is not too large. 

Using these approaches, we calculate the SEDs of forming giant planets over a wider range of parameter space, including variations in the mass, the mass accretion rate, the geometry of the infall at the outer boundary, the semimajor axis of the orbit, the opacity properties of the system, and the viewing angle. We also consider effects due to the background circumstellar disk. We also present synthetic images of the forming planets. We find observable differences in the SEDs and in images of the planet/disk/envelope system based on the envelope density (infall) distribution, allowing future observations to distinguish between these models. The speed of our model also allows for efficient model fitting to future observations.


This paper is structured as follows --- Sec. \ref{sec:priorres} discusses the structure of the circumplanetary environment and introduces our formulation for the structure of the forming planet. Sec. \ref{sec:SEDs} presents our methodology for calculating the spectral energy distributions of the system, including a fully numerical treatment and a parametric approximation. Sec. \ref{sec:SEDresults} explores parameter space presenting SEDs for a range of system parameters. In Sec. \ref{sec:images}, we calculate synthetic images of the planet/disk/envelope system for comparison with future spatially-resolved observations. Sec. \ref{sec:backgrounddisk} discusses how the background circumstellar disk can affect the radiative signature of the forming planet. Finally, we conclude in Sec. \ref{sec:disc} with a summary of our results and a discussion of their implications. 

\section{Protoplanet Structure}\label{sec:priorres}

In this section, we specify the structure of the forming giant planet. During its formation stages, the planet itself is surrounded by a circumplanetary disk, which is embedded within an envelope of accreting material. The SEDs of forming planets are determined by the radiation generated by the central planet and disk and is processed through the envelope. 

The central planet/disk system is fed by a collapse flow that originates at the boundary of the sphere of influence of the planet. Beyond the boundary, the background circumstellar disk provides the incoming material. The boundary is nominally given by the Hill radius $R_H=a(M/3M_\star)^{1/3}$, where $a$ is the semimajor axis of the planetary orbit and $M$ is the planet mass. However, if the Bondi radius $R_B=GM/c_s^2$ is smaller than the Hill radius ($c_s$ is the local sound speed), then the Bondi radius will mark the outer boundary of the system. Further complications arise if the circumstellar disk scale height $H=c_s/\Omega$ is smaller than the nominal outer boundary, as the background disk density gradient modifies the conditions at the outer boundary of the forming planet. In general, however, these three scales are roughly comparable. The Hill radius is the effective boundary in most instances, although the finite size of the scale height $H$ plays a role when the planet is high-mass. In this paper, we use the Hill radius as the nominal boundary scale, and these results are easily modified for other conditions. Throughout this text, `Hill radius' refers to the generic outer boundary. 

In order to specify the structure of the system, we must set the mass inflow rate at the outer boundary. At a given time, the circumstellar disk provides mass accretion into the Hill sphere at a rate ${\dot M}$. This quantity represents the net accretion rate, as some of the incoming material flows back out of the Hill sphere before reaching the central regions (e.g., \citealt{Lambrechts2017,Lambrechts2019}). In addition, the distribution of material entering the sphere of influence of the planet is not necessarily isotropic. To account for different possible inflow geometries, we introduce asymmetry functions $f_i$ such that the mass inflow rate at the outer boundary is given by $\dot{M}f_i(\mu_0)$, where $\dot{M}$ is the total net mass inflow rate and $\mu_0=\cos\theta_0$ is the cosine of the initial polar angle, measured with respect to the planetary pole and evaluated at the outer boundary.  These functions are chosen to have the form 
\begin{equation}\label{eq:asyminfunc}
    f_i(\mu_0)=\begin{dcases}
        3\mu_0^2 & \text{polar;}\\
        2\mu_0 & \text{quasipolar;}\\
        1 & \text{isotropic;} \\
        \frac{4}{\pi}(1-\mu_0)^{1/2} & \text{quasiequatorial;} \\
        \frac{3}{2}(1-\mu_0) & \text{equatorial.}
    \end{dcases}
\end{equation}
The labels are introduced for convenience. The functions $f_i$ are normalized such that 
\begin{equation*}
    \int\displaylimits_0^1\!\!f_i(\mu_0)\,\de\mu_0=1\,.
\end{equation*}

As material falls from the outer boundary toward the central planet/disk system, we assume that gravitational forces dominate over pressure forces such that the flow becomes ballistic (see \citealt{Adams2022} for an accounting of this approximation and others; see also \citealt{Mendoza2009} for a generalization). This scenario thus requires that the incoming gas can sufficiently cool.\footnote{In the absence of cooling, the incoming gas heats up, pressure forces increase, and planet formation can be compromised. We work in the regime where cooling is efficient and giant planet formation can be successful. }  In this case, incoming parcels of gas execute zero-energy Keplerian orbits about the central planet and obey an orbit equation of the form \citep{Ulrich1976}
\begin{equation}\label{eq:orbit}
    1-\frac{\mu}{\mu_0}=\zeta(1-\mu_0^2)\,.
\end{equation}
In this equation, the parameter $\zeta\equiv R_C/r$, where $R_C$ is the centrifugal radius defined below. The orbit equation (\ref{eq:orbit}) can be used to solve for $\mu_0$ for a given location ($\mu$, $r$). This procedure also specifies the velocity field of the incoming flow \citep{Ulrich1976,Cassen1981,Chevalier1983}. In particular, the radial velocity of the incoming material is given by 
\begin{equation}\label{eq:vr}
    v_r=-\left[\frac{GM}{r}\left[2-\zeta(1-\mu_0^2)\right]\right]^{1/2}\,.
\end{equation}
Using conservation of mass along streamlines, the density in the envelope can be written as 
\begin{equation}\label{eq:envrho}
    \rho(r, \mu)=\frac{\dot{M}f_i(\mu_0)}{4\pi r^2|v_r|}\big[1+\zeta(3\mu_0^2-1)\big]^{-1}\,.
\end{equation}
Note that $f_i(\mu_0)$ is the inflow asymmetry function given by Eq.~\eqref{eq:asyminfunc} and that the cosine $\mu_0$ of the initial polar angle can be determined from the position ($r, \mu$) using Eq. \eqref{eq:orbit}.

The circumplanetary disk that forms is bounded on the inside by the magnetic truncation radius and (initially) on the outside by the centrifugal radius, where material with the highest angular momentum will intersect the equatorial plane. The magnetic truncation radius takes the form
\begin{equation}\label{eq:RXdef}
    R_X=\omega\left[\frac{B_p^4 R_p^{12}}{GM\dot{M}^2}\right]^{1/7}\,.
\end{equation}
In Eq.~\eqref{eq:RXdef}, $\omega$ is a dimensionless constant of order unity {(here we assume $\omega=1$)}, $B_p$ is the planetary magnetic field, $R_p$ is the radius of the planet, $M$ is the mass of the central planet, and $\dot{M}$ is the mass accretion rate (e.g., see \citealt{Ghosh1978, Blanford1982, Lovelace2011}). In practice, the magnetic truncation radius is several times the planet radius, which is itself about twice the radius of Jupiter \citep{Adams2022}. {For a fiducial model with parameters given in Table \ref{tab:canonvals}, the truncation radius $R_X\simeq3.8R_p$.} The nominal outer boundary of the disk is given by the centrifugal radius of the collapse flow (see \citealt{Quillen1998, Tanigawa2012}) and is given by 
\begin{equation}\label{eq:RCdef}
    R_C=\frac{1}{3}R_H=\frac{a}{3}\bigg(\frac{M}{3M_\star}\bigg)^{1/3}\,.
\end{equation}

The steady-state surface density of the disk can also be calculated from the source function given by the infall solution and standard disk evolution. In the presence of sufficient viscosity, the surface density generally has a power-law form that depends on the temperature profile and the viscosity \citep{Hartmann2009}, where modest variations from this form depend on the infall geometry.\footnote{{In this case, the surface density $\Sigma\propto r^{-1}$. However, our models of the SEDs assume that the disk is optically thick and are otherwise independent of the surface density.}} For cases where giant planet formation is successful, the circumplanetary disk must process sufficient amounts of material in order to form the planet. The viscosity must be large enough to drive disk accretion on the infall time scale, so that the disk surface density approaches the steady-state form. The required viscosity is of order $\alpha\sim\num{e-4}$, significantly smaller than that required in the star formation problem (see \citealt{Adams2022} for a discussion of this value). We also assume that the circumplanetary disk is optically thick to its own emitted radiation, which is the case for typical parameters. 

Finally, we calculate the luminosity of the planet and the disk. The planet luminosity is given by 
\begin{equation}\label{eq:Lp}
    L_p=\frac{GM\dot{M}}{R_p}\Bigg[1-\frac{R_p^3}{3R_X^3}\Bigg]\Bigg[1-f_d\frac{R_p}{R_X}\Bigg]\,.
\end{equation}
In Eq.~\eqref{eq:Lp}, we have assumed that the planet's internal luminosity is small compared to the energy released by infalling mass (see \citealt{Marley2007}). The first term in Eq. \eqref{eq:Lp} is the nominal energy delivered to the system if all of the infalling material struck the planet. The second term accounts for the rotational energy of the planet, assuming that the planet corotates with the inner edge of the disk. The third term accounts for the relative fraction of material that directly falls onto the planet versus the fraction that is processed through the disk. The fraction $f_d$ is the fraction of infalling material that initially strikes the disk, which is given by 
\begin{equation}\label{eq:approxints}
    f_d\simeq\begin{dcases}
        1-3u_p/2&\text{polar;}\\
        1-u_p&\text{quasipolar;}\\
        1-u_p/2&\text{isotropic;}\\
        1-(4/3\pi)u_p^{3/2}&\text{quasiequatorial;}\\
        1-(3/8)u_p^2&\text{equatorial},
    \end{dcases}
\end{equation}
where we have defined $u_p=R_p/R_C$. Assuming that the planet emits as a blackbody, its temperature is given by 
\begin{equation}\label{eq:Tp}
    T_p=\left(\frac{L_p}{4\pi\sigma R_p^2}\right)^{1/4}\,.
\end{equation}
The luminosity of the disk takes the form 
\begin{equation}\label{eq:Ld}
    L_d=f_d\frac{GM\dot{M}}{2R_X}\,.
\end{equation}
Eq. \eqref{eq:Ld} assumes that the material processed through the disk falls through a gravitational potential well with depth given by $M/R_X$ and that half of the energy is stored as rotational energy. The irradiation of the disk by the planet is assumed to be small.\footnote{As shown in later sections, this approximation is valid for the regimes of parameter space under consideration in this paper. Analytic calculations also show that the luminosity absorbed by the disk beyond the truncation radius is a small fraction of the planet's luminosity, which is comparable to the disk accretion luminosity.} We have again assumed that the disk viscosity is large enough that most of the material landing on the disk accretes onto the planet \citep{Zhu2015, Szulagyi2017, Zhu2018}. We then set the disk temperature to follow
\begin{equation}\label{eq:Td}
    T_d(r)=T_X\!\left(\frac{r}{R_X}\right)^{-3/4}\,.
\end{equation}
The constraint on the total luminosity of the disk means that we require
\begin{equation}\label{eq:TX}
    \sigma T_X^4=f_d\frac{GM\dot{M}}{8\pi R_X^3}\left[1-\frac{R_X}{R_C}\right]^{-1}\,,
\end{equation}
assuming that the disk is optically thick.

As long as the viscosity is large enough so that the circumplanetary disk reaches steady-state, the disk structure is largely independent of the geometry of the inflow. The differences in the fractions of material accreted directly onto the planet and processed through the disk vary only by factors of order unity. As a result, the most  significant changes due to the different inflow functions manifest in the envelope. Notice that the velocity of material in the envelope has no effect on the radiative signatures --- only the density distribution is relevant. As a result, although we calculate the radiation fields based on the inflow geometry, we are actually investigating the response of the radiative signatures to the envelope density distribution. These results apply not only to ballistic trajectories (as assumed in our calculations) but also any circumplanetary envelopes with analogous density distributions. 

\setlength{\tabcolsep}{2pt}
\begin{table}[t]
    \caption{\textbf{Canonical Fiduciary Values.} Fiduciary values for model parameters, which are repeatedly used throughout this paper. The opacity is assumed to follow $\kappa_\nu=\kappa_0(\nu/\nu_0)^\eta$.}
    \centering
    \begin{tabular}{c|rl}
         \multicolumn{3}{c}{System Properties}\\\hline
         $M_\star$ & 1 & \unit{M_\odot} \\
         $a$ & 5 & au \\
         $\dot{M}$ & 1 & \unit{M_J\per\mega yr} \\
    \end{tabular}
    \begin{tabular}{c|rl}
         \multicolumn{3}{c}{Opacity Properties}\\\hline
         $\kappa_0$ & 10 & cm$^2$ g$^{-1}$ \\
         $\nu_0$ & $10^{14}$ & Hz \\
         $\eta$ & 1 & \\
    \end{tabular}
    \begin{tabular}{c|rl}
         \multicolumn{3}{c}{Planet Properties}\\\hline
         $M$ & 1 & M$_{\rm J}$ \\
         $R_p$ & $10^{10}$ & cm \\
         $B_p$ & 500 & G \\
    \end{tabular}
    \label{tab:canonvals}
\end{table}

\begin{figure*}[t!]
    \centering
    \includegraphics[width=\linewidth]{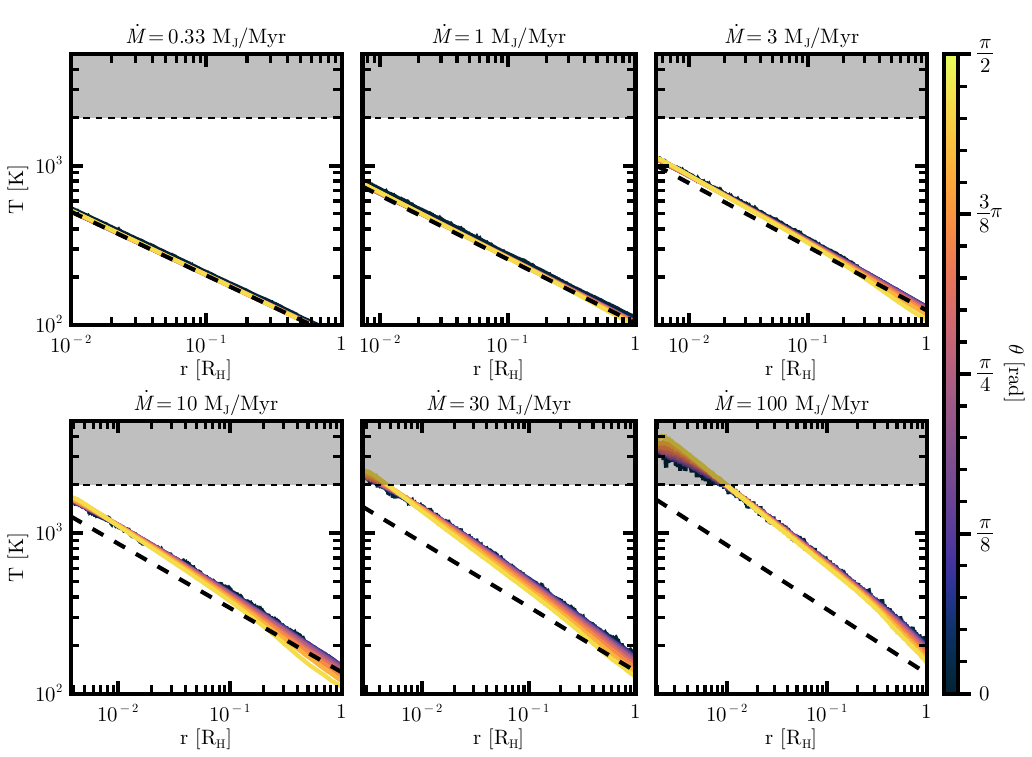}
    \vspace{-15pt}
    \caption{The temperature distribution of the circumplanetary envelope for a range of mass inflow rates, comparing \texttt{RADMC-3D} and the parametric temperature approximation. Results are shown for the isotropic inflow function (although the behavior is similar for the other inflow geometries). The dashed black curve depicts the parametric approximation, while the colored curves show the temperature at different angles throughout the envelope. The envelope is axisymmetric and mirrored across the equatorial plane. The line through $\theta=\pi/2$ is not shown, since this ray corresponds to the circumplanetary disk midplane. The left-hand boundary of each panel is the magnetic truncation radius $R_X$. Although it is not accounted for, the envelope's dust opacity will vanish at $T\gtrsim\qty{2000}{K}$, which is marked by a shaded region.}
    \vspace{-15pt}
    \label{fig:tempgrids}
\end{figure*}

\begin{figure*}[t!]
    \centering
    \includegraphics[width=\linewidth]{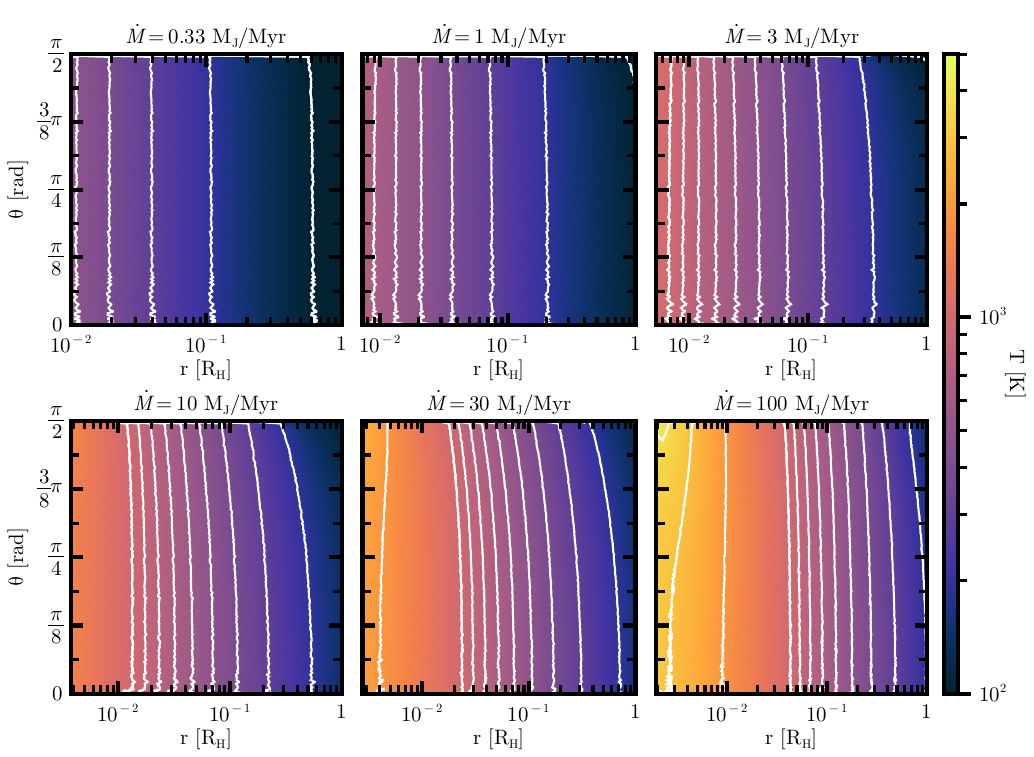}
    \caption{{The temperature distribution in the $(r, \theta)$ plane of the circumplanetary envelope for a range of mass inflow rates. Results are shown for models calculated using \texttt{RADMC-3D}. Contours are shown at every minor tick on the colorbar (100, 200,..., 1000, 2000, ...). The small oscillations in the contours are the result of numerical noise in the temperature model. The direction $\theta=0$ corresponds to the planetary pole. Note that the contours show relatively little angular dependence. }}
    \vspace{-15pt}
    \label{fig:rthetatempgrid}
\end{figure*}

\begin{figure*}[t!]
    \centering
    \includegraphics[width=\linewidth]{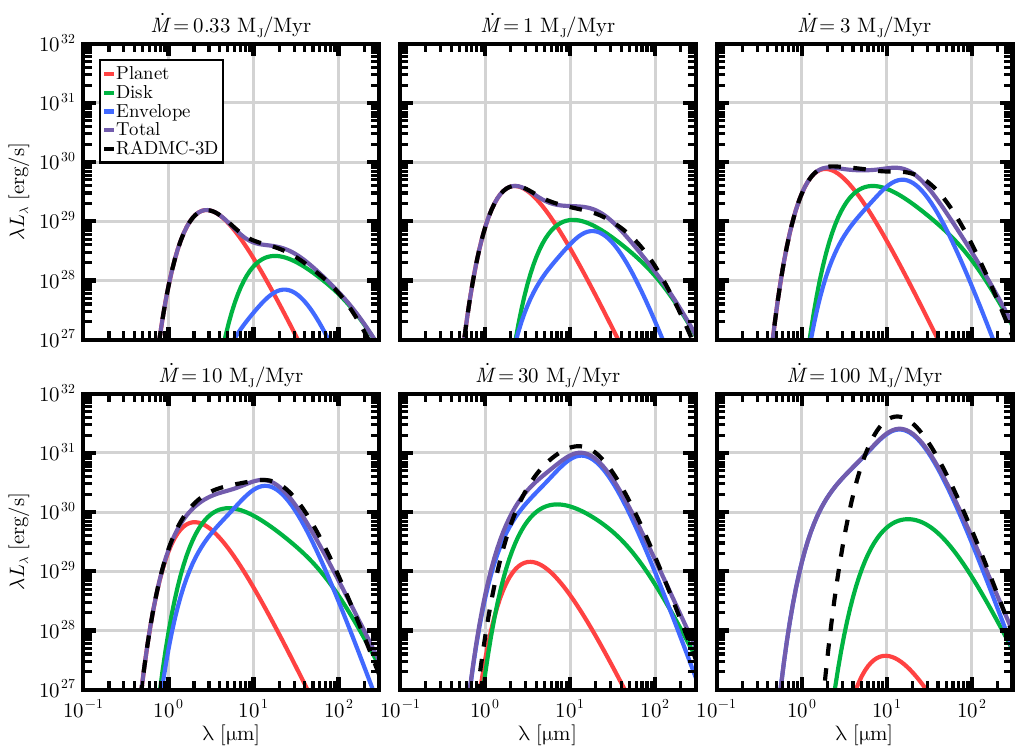}
    \vspace{-15pt}
    \caption{Comparison of the spectral energy distributions calculated from the parametric approach and the numerical package \texttt{RADMC-3D}. The planet and disk SEDs are attenuated by the envelope, which then reemits the absorbed luminosity. The system is assumed to have isotropic inflow at the outer boundary and is viewed from above ($\psi=0$). Varying the inflow geometry or the viewing angle does not significantly modify the comparison. Aside from the mass inflow rate, the parameters are given in Table~\ref{tab:canonvals}. }
    \vspace{-15pt}
    \label{fig:SEDMCcomp}
\end{figure*}

\section{Spectral Energy Distributions}\label{sec:SEDs}

In this section, we discuss our methodology for computing predicted SEDs for the system. This approach generalizes previous efforts by allowing the disk to have physical extent and takes into account the asphericity of the envelope. 

In order to calculate the SED, we consider the case where the observation point --- a detector --- resides at infinity. The detector plane is a circle with a radius $R_H$, offset from the polar direction by a viewing angle $\psi$ and tangent to the Hill surface. The wavelength-dependent luminosity incident on this detector is the integral of the flux at each point over its area. If the detector is at infinity, however, each point on the detector only receives intensity from the ray normal to the detector surface. We calculate the {intensity} at each point with a ray-tracing method {and calculate the equivalent spherical luminosity by multiplying by $4\pi$ steradians. For the remainder of the text, ``luminosity'' refers to the equivalent spherical luminosity}. For ease of integration, we ray-trace from all points in the system to the detector plane, rather than from the detector to the system. This {approach} significantly improves the integration time and accuracy. We also break the intensity down into the contributions of the three parts of the system --- the planet, the disk, and the envelope --- and solve each separately. \

\subsection{Planet Spectral Luminosity}\label{subsec:planetSED}

We first consider the luminosity from the planet. Assuming that the planet is sufficiently small that the column density between the detector plane and the planet is a constant across the planet-striking rays, the total luminosity due to the planet is
\begin{equation}\label{eq:Lpnu}
    L_{p,\nu}=4\pi\,\pi R_p^2\, B_\nu(T_p)\,\exp(-\kappa_\nu N_{\rm col})\,.
\end{equation}
In Eq.~\eqref{eq:Lpnu}, $B_\nu$ is the blackbody function, $R_p$ is the planet's radius, $\kappa_\nu$ is the opacity, $N_{\rm col}$ is the column density, and $T_p$ is the planet's temperature, as given by Eq.~\eqref{eq:Lp} under the assumption of blackbody emission. The leading factor of $4\pi$ comes from the {definition of the equivalent spherical luminosity}, and the factor of $\pi R_p^2$ is the projected area of the planet in the detector plane. The column density can be calculated by integrating the density (Eq. \ref{eq:envrho}) from ($R_H,\cos\psi$) to ($R_p,\cos\psi$) along the ray $\de r$. 

\subsection{Disk Spectral Luminosity}\label{subsec:diskSED}

The luminosity due to the disk is slightly more complicated. Consider a point on the disk given by ($r',\phi$), where $r'$ is the distance from the origin in the disk plane and $\phi$ is the azimuthal angle relative to the detector direction. We calculate the luminosity that this point provides to the detector plane along the normal ray. The most complicated factor in this method is the calculation of the column density between these points. We assume for convenience that the density of the environment outside of the Hill sphere is zero, although the system is in fact embedded in the background protoplanetary disk (this issue is addressed in Sec. \ref{sec:backgrounddisk}). 

We define a Cartesian coordinate system to calculate the density along the viewing ray.  The $z$ direction is chosen to lie along the rotational pole of the planet. We set the $x$-direction to be pointed at the detector and restricted to the equatorial plane. The $y$-direction is chosen to ensure that the coordinate frame is right-handed. The disk point is given by $\Vec{p}_d=[r'\cos\phi,\,r'\sin\phi,\,0]$, and the unit vector pointing along the ray is given by $\hat{n}=[\sin\psi,\,0,\,\cos\psi]$. At a distance $s$ along the ray, the position is given by $\Vec{p}_d+s\hat{n}$. This point has radius $r$ and polar parameter $\mu$ given by
\begin{subequations}\label{eq:diskrays}
\begin{align}
    r&=\sqrt{r'^2+s^2+2r's\sin\psi\cos\phi}\,;\\
    \mu&=\frac{s}{r}\cos\psi\,.
\end{align}     
\end{subequations}
The column density between the {disk} point and the detector plane is given by 
\begin{equation}
    N_{\rm col, d} = \int\displaylimits_0^{s_0}\!\!\de s\, \rho(r[s], \mu[s])\,. \label{eq:dNcol}
\end{equation}
In Eq.~\eqref{eq:dNcol}, $s_0$ is the value of $s$ such that $r(s_0)=R_H$, where $r(s)$ and $\mu(s)$ are given by Eq. \eqref{eq:diskrays}. 

Since the disk is a blackbody emitter at each point, the total intensity on the detector plane is given by the (attenuated) blackbody intensity. Instead of integrating over the detector plane, we integrate over the disk and introduce a factor of $\cos\psi$ to account for the projection of the disk area. The total luminosity of the disk is then
\begin{equation}\label{eq:Ldnu}
    L_{d,\nu}=4\pi\cos\psi\!\!\int\displaylimits_0^{2\pi}\!\!\de\phi\!\!\int\displaylimits_{R_X}^{R_C}\!\!\de r\, r B_\nu(T_d)\,\exp(-\kappa_\nu N_{\rm col,d})\,.
\end{equation}

\subsection{Envelope Spectral Luminosity}\label{subsec:envL}

Next we consider the luminosity due to the envelope. In order to generate SEDs, we must accurately calculate the radiative transfer through the envelope. We consider two different methodologies for the radiative transfer. First, we assume that the temperature of the envelope is spherically symmetric and optically thin to its own radiation, which sets the form of the envelope's temperature distribution. The parameters are then set by ensuring conservation of luminosity. Second, we use a radiative transfer Monte Carlo to calculate the envelope's temperature and spectral energy distribution. These two methodologies are complementary --- the parametric approximation is significantly more efficient than the fully numerical method, but is only accurate in a restricted range of parameter space. 

\subsubsection{Parametric Temperature Approximation}

In the case of an optically thin envelope, the temperature distribution will be nearly spherically symmetric. The form of the temperature distribution will be similar to a spherically symmetric cloud illuminated by a central point source. Although the disk is finite in extent, reaches a maximum size of $1/3$ of the envelope radius. In addition, the majority of the disk luminosity originates from its inner regions. As a result, the finite disk will still generally behave like a point source. In the optically thin limit, the aspherical density distribution of the envelope will be a similarly minor factor. We can then use the spherical temperature distribution calculated by \citet{Adams1985}, which takes the form
\begin{equation}
    T_e(r)=T_C\left(\frac{r}{R_C}\right)^{-2/5}\,. \label{eq:Tesphere}
\end{equation}

In Eq. \eqref{eq:Tesphere}, the constant $T_C$ is set so that the total luminosity of the system is constant. This form assumes a linear opacity law ($\eta=1$), as changes in the opacity index modify the index of the temperature distribution (see Sec. \ref{subsec:kappacomp}). The total luminosity of the envelope is given by the integral over both the volume and over all frequencies, 
\begin{equation}
    L_e(r) = \int_V dV \int\displaylimits_0^\infty\!\!\de\nu\,4\pi\rho\kappa_\nu B_\nu(T_e) \,, 
\end{equation}
where $\sigma$ is the Stefan-Boltzmann constant and $\kappa_\nu$ is the frequency-dependent opacity. This expression can be evaluated to find that 
\begin{equation}
    L_e = 16\pi R_C^2 b_\kappa \sigma T_C^5 N_{\rm col}\,,\label{eq:Letotal}
\end{equation}
where a linear opacity law is used and $b_\kappa$ is the coefficient of the Planck mean opacity (see Appendix \ref{sec:bolokappa}). In Eq. \eqref{eq:Letotal}, we have expressed the integral of the density over the envelope in terms of the average column density $N_{\rm col}$, which can be calculated either analytically (see \citealt{Taylor2024}) or numerically from Eq. \eqref{eq:envrho}.  

In order to enforce energy conservation, we require that the total energy emitted by the envelope $L_e$ is equal to the total energy absorbed by the envelope. The total energy absorbed by the system is the total energy delivered $L_p+L_d$ minus the energy that escapes. In Secs. \ref{subsec:planetSED} and \ref{subsec:diskSED}, we calculated the specific luminosities that escape the system at a given viewing angle $\psi$. To calculate the total energy released, we integrate over the frequency $\de\nu$ and over the entire Hill sphere. The integral over the Hill sphere is simplified by the fact that the system is azimuthally symmetric, so that the envelope luminosity is given by  
\begin{equation}
    L_e = L_p+L_d-4\pi\!\!\int\displaylimits_0^1\!\!\de\mu\!\!\int\displaylimits_0^\infty\!\!\de\nu\,\left[L_{p,\nu}+L_{d,\nu}\right]\,. \label{eq:Lebalance}
\end{equation}
For convenience, the integrals $\de\nu$ are evaluated analytically (see Sec. \ref{sec:images} and Appendix \ref{subsec:alphaderiv}). Combining Eqs. \eqref{eq:Letotal} and \eqref{eq:Lebalance}, we solve for the temperature constant $T_C$ and the envelope's temperature distribution.

Given the temperature distribution of the envelope, we calculate the emitted specific luminosity. Since we have assumed that the envelope is optically thin to its own radiation, there is no attenuation. A single point at temperature $T$ emits intensity $\rho\kappa_\nu B_\nu(T)$ in all directions. Since the radiation is not attenuated, the total intensity seen at a viewing angle $\psi$ is the integral of the point intensities over the entire envelope. Therefore, the total luminosity due to the envelope is 
\begin{equation}
    L_{e, \nu}=16\pi^2\!\!\int\displaylimits_0^1\!\!\de\mu\!\!\int\displaylimits_0^{R_H}\!\!\de r\,r^2\rho(r,\mu)\kappa_\nu B_\nu(T_e[r])\,.\label{eq:Lenu}
\end{equation}
In Eq. \eqref{eq:Lenu}, there is a factor of $4\pi$ from the integral over the envelope and a factor of $4\pi$ {steradians from the definition of the equivalent spherical luminosity}.

\subsubsection{Radiative Transfer Monte Carlo}

An alternate method is to solve the radiative transfer problem via a Monte Carlo methodology. We used the open-source software \texttt{RADMC-3D} \citep{Dullemond2012} to model the temperature distribution of the envelope {and} to generate the SEDs. 

We use a 2D, axisymmetric spherical domain set to the size of the Hill sphere. The radial axis is divided into 1000 log-spaced cells and the polar axis is divided into 150 evenly-spaced cells from $\theta=0$ to $\theta=\pi/2$. The domain is mirrored across the equatorial plane. We place a blackbody planet in the center, set to radiate the luminosity given by Eq. \eqref{eq:Lp}. For the disk, we use \texttt{RADMC-3D}'s internal heating functionality. The single row of grid cells closest to the disk midplane are given an internal heat source equal to the total luminosity radiated by the disk over the same radius range. The density of these disk cells are set such that they have a Planck-mean optical depth of $\kappa_P N_{\rm col}=1$. This ensures that the temperature of the disk cells is equivalent to the expected temperature of the thin disk (given by Eqs. \ref{eq:Td} and \ref{eq:TX}). The density of the rest of the envelope is given by Eq. \eqref{eq:envrho}. The system opacity is set to follow a power law such that 
\begin{equation}
    \kappa_\nu=\kappa_0\left(\frac{\nu}{\nu_0}\right)^\eta\,,
\end{equation}
and we set $\eta=1$ (see Sec. \ref{subsec:kappacomp} for a discussion of this value). We then use \texttt{RADMC-3D} to calculate the temperature at every point in the envelope, using $10^6$ photon packets.

\subsubsection{Methodology Comparison}

In Fig. \ref{fig:tempgrids}, we compare the temperature distributions calculated by \texttt{RADMC-3D} to the spherically symmetric parametric temperature distribution given by Eq. \eqref{eq:Tesphere} for a range of mass inflow rates. All other parameters are kept fixed at the fiducial values given in Table \ref{tab:canonvals}. {In Fig. \ref{fig:rthetatempgrid}, we plot the temperature distribution from the numerical package \texttt{RADMC-3D} in the $(r,\theta)$ plane.} Although we only show these comparisons for the isotropic inflow geometry, the other geometries have similar behavior. As shown in the figure, the parametric approximation accurately determines the temperature distributions and SEDs for the majority of the mass inflow rates, and {only} departs from the Monte Carlo results for mass infall rates $\dot{M}\gtrsim\qty{30}{M_J/\mega yr}$. The discrepancy at high mass inflow rates results from the high optical depth of the envelope,\footnote{Varying the mass inflow rate and the planet mass to keep the optical depth constant reveals that this divergence is largely due to the optical depth.} which results in a more complicated temperature profile in the inner portion of the envelope. However, the parametric approximation represents a significant improvement in evaluation time. 

Eqs.~\eqref{eq:Lpnu}, \eqref{eq:Ldnu}, and \eqref{eq:Lenu} all depend on the mass inflow rate $\dot{M}$ (via the density), the inflow function, and the viewing angle $\psi$. For comparison, in Fig. \ref{fig:SEDMCcomp} we show the SEDs that result from both of our methodologies. Although we only show comparisons for isotropic inflow at one viewing angle, varying these parameters cause little difference in the qualitative fidelity of the comparison. As expected from the results of Fig. \ref{fig:tempgrids}, the two models are generally similar, although they diverge at sufficiently high optical depths. In this limit of large optical depth, the envelope reprocesses most of the radiation from the central planet and disk, so that the resulting SEDs approach a blackbody-like form (see the final panel of Fig. \ref{fig:SEDMCcomp}).  

For this comparison, we assume that the opacity coefficient $\kappa_0$ for the circumplanetary dust has a value of \qty{10}{\centi\meter^2\per\gram}, which is typical of the interstellar medium \citep{Semenov2003}. As the planet formation process advances, the opacity is expected to decrease significantly as the dust settles and coagulates (e.g., \citealt{Weidenschilling1984, Dominik1997, Dullemond2005}). As a result, we expect that the parametric approximation will become increasingly applicable as the system evolves and the optical depth decreases. Further consequences of opacity variations are addressed in Sec. \ref{subsec:kappacomp}. 

\begin{figure*}
    \centering
    \vspace{-5pt}
    \includegraphics[width=\linewidth]{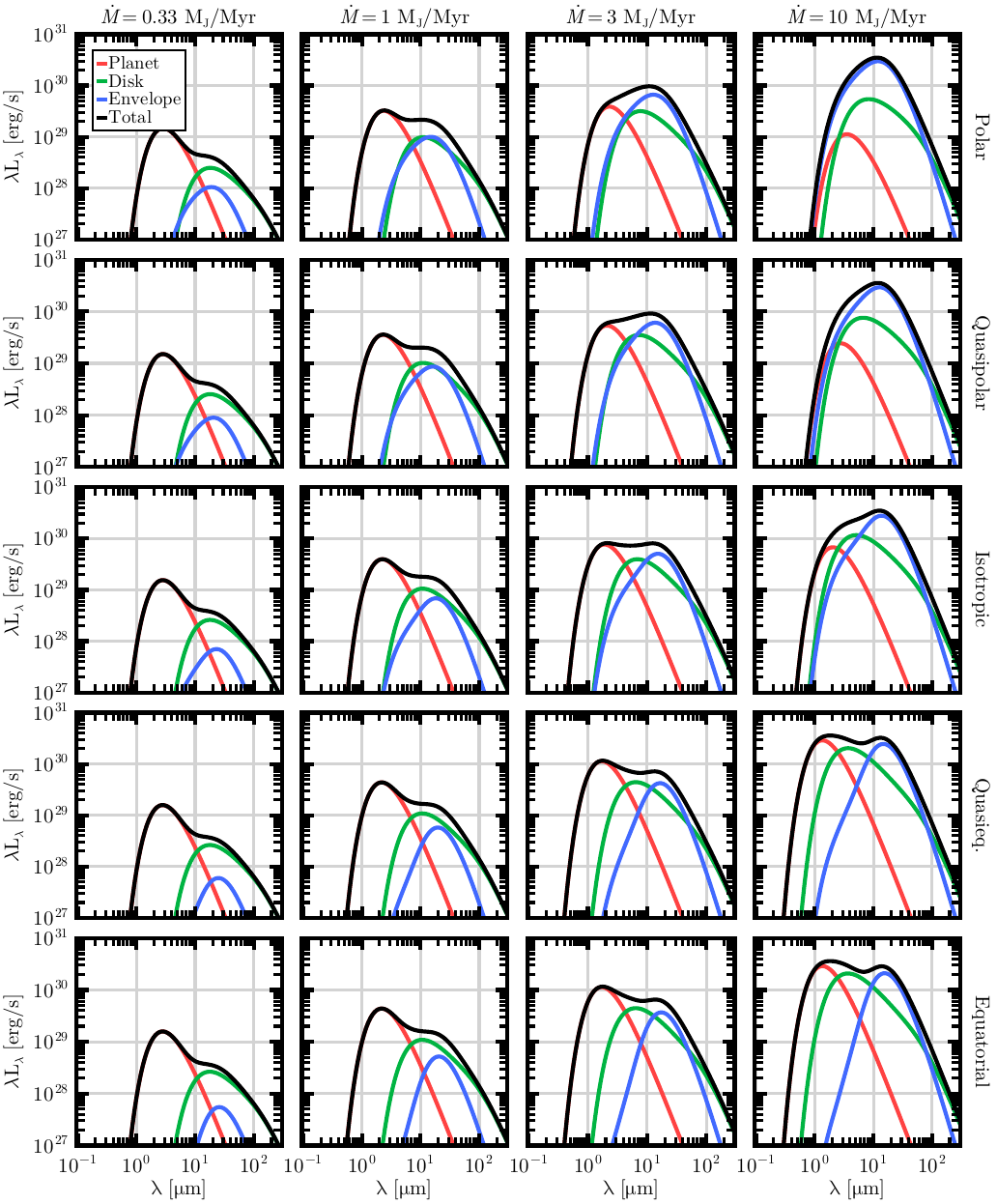}
    \vspace{-15pt}
    \caption{The spectral energy distributions of the planet/disk/envelope system for a range of mass inflow rates. The system is assumed to be viewed from above. Aside from the mass inflow rate, the system parameters are set to their fiducial values (Table~\ref{tab:canonvals}).}
    \vspace{-15pt}
    \label{fig:SEDgridmdot}
\end{figure*}

\begin{figure*}[t!]
    \centering
    \includegraphics[width=\linewidth]{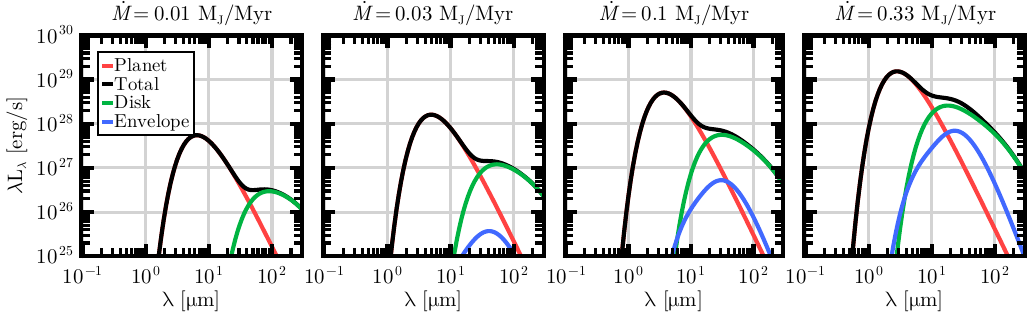}
    \vspace{-15pt}
    \caption{{The spectral energy distributions of the planet/disk/envelope system for substantially lower values of the infall rate $\dot{M}$. These lower rates are closer to those measured for the known protoplanets PDS 70 b/c and can be realized in the later stages of infall. }}
    \vspace{-15pt}
    \label{fig:SEDgridlowMdot}
\end{figure*}

\begin{figure*}
    \centering
    \vspace{-5pt}
    \includegraphics[width=\linewidth]{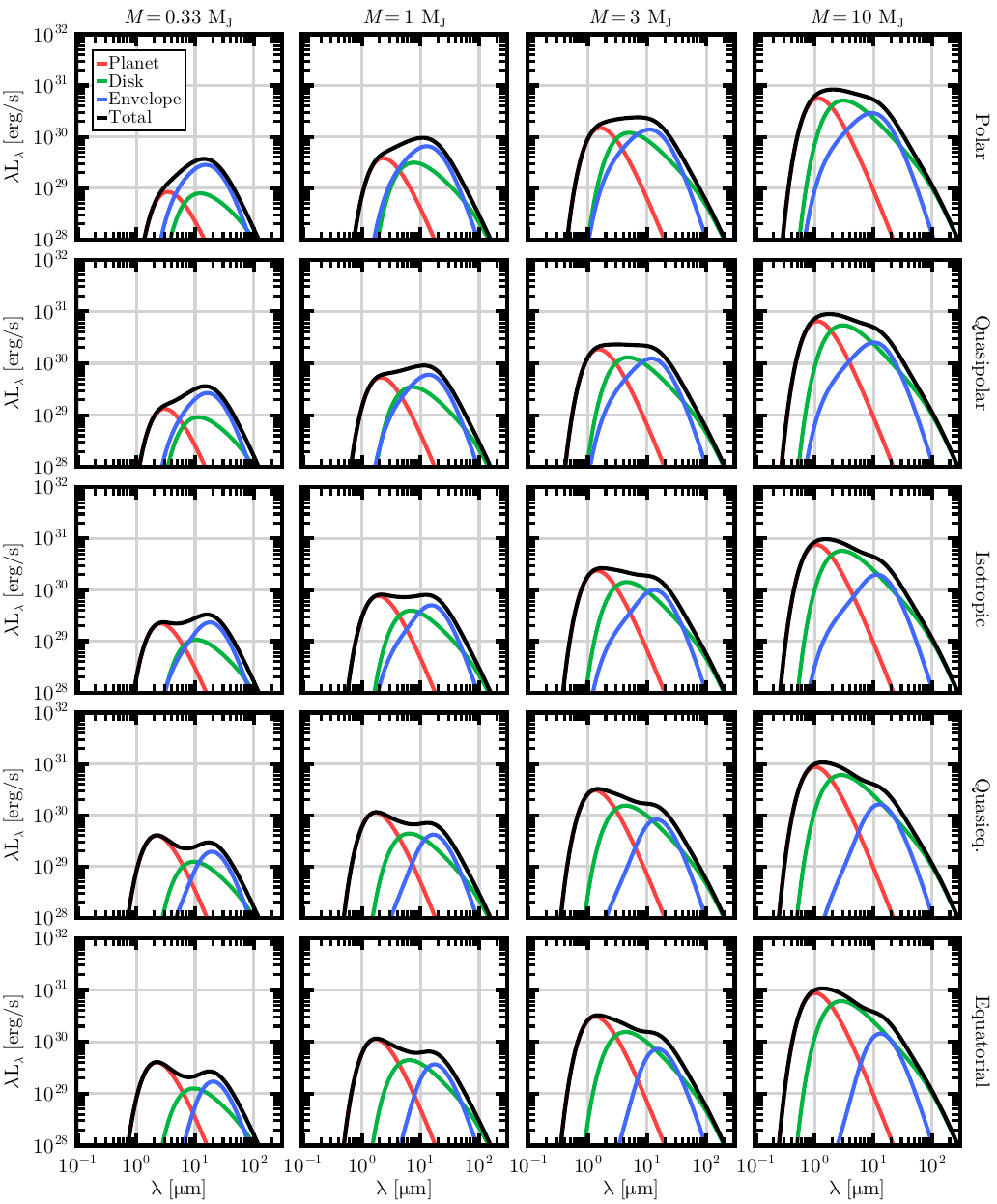}
    \vspace{-15pt}
    \caption{The spectral energy distributions of the planet/disk/envelope system for a range of planet masses. The system is assumed to be viewed from above. Aside from the planet mass, the system parameters are set to their fiducial values (Table~\ref{tab:canonvals}).}
    \vspace{-15pt}
    \label{fig:SEDgridm}
\end{figure*}

\begin{figure*}
    \centering
    \vspace{-20pt}
    \includegraphics[width=\linewidth]{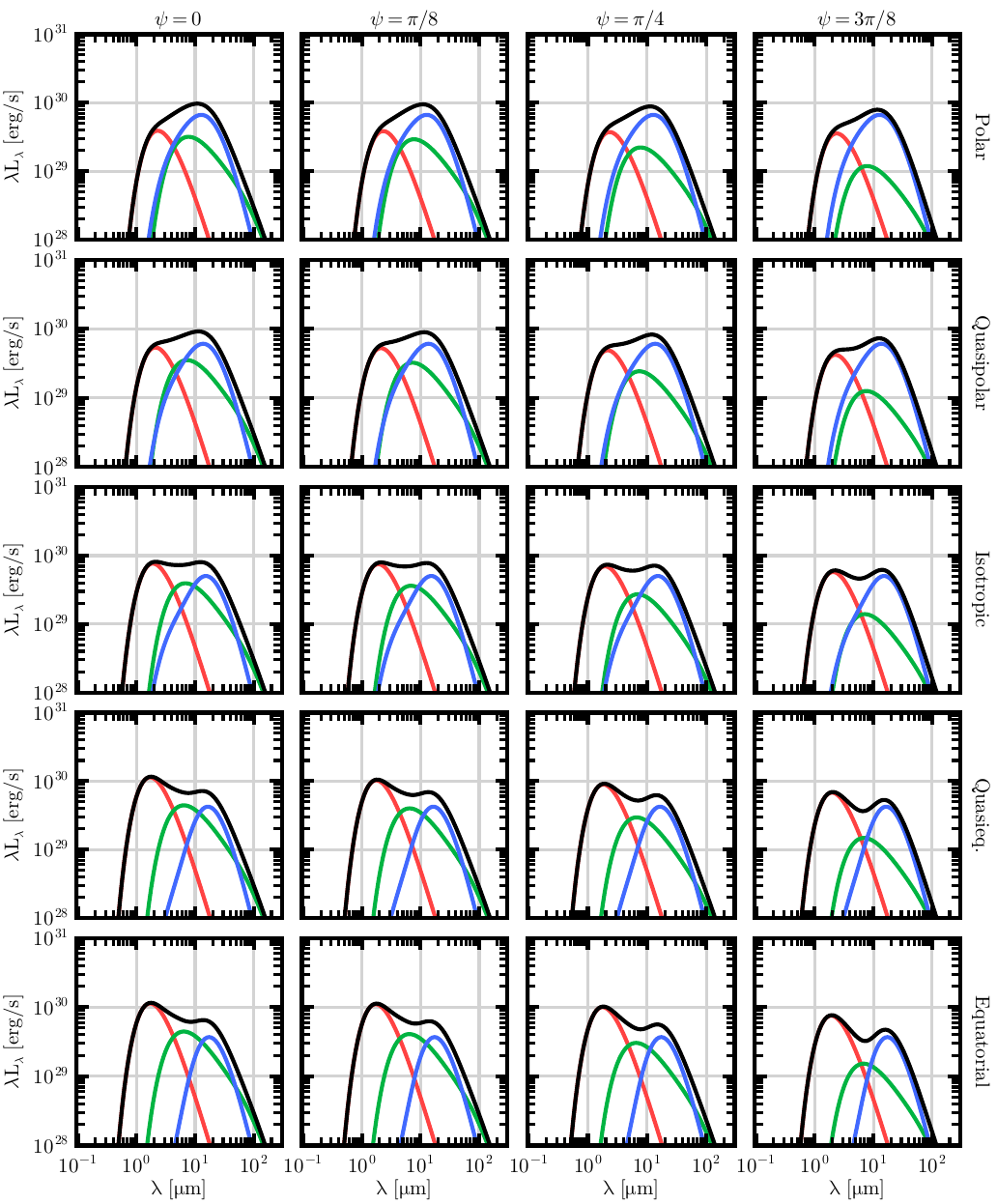}   
    \vspace{-15pt}
    \caption{The spectral energy distributions of the planet/disk/envelope system for a range of viewing angles. The mass inflow rate is assumed to be $\dot{M}=3$ $M_J$/Myr. The red, green, and blue curves depict the planet, disk, and envelope SEDs, respectively. The black curve shows the total SED of the planet/disk/envelope system. All system parameters are set to their fiducial values (Table~\ref{tab:canonvals}).}
    \vspace{-20pt}
    \label{fig:SEDgridangle}
\end{figure*}

\begin{figure*}[t!]
    \centering
    \includegraphics[width=\linewidth]{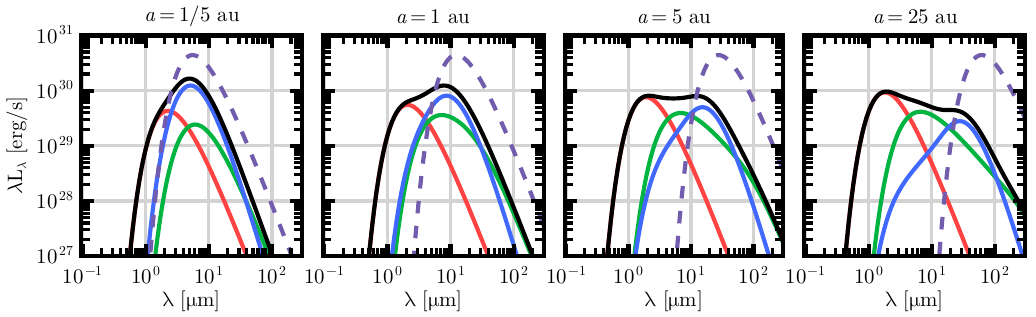}
    \vspace{-15pt}
    \caption{The spectral energy distributions of the planet/disk/envelope system for a range of semimajor axes. The system is assumed to be viewed from above. Aside from the semimajor axis, the parameters are given in Table~\ref{tab:canonvals}. The red, green, and blue curves depict the contributions from the planet, disk, and envelope SEDs, respectively. The black curve shows the total SED of the forming planet. For comparison, the purple dashed curve shows the SED emitted by a portion of the background circumstellar disk --- a surface with the area of the Hill sphere, the temperature of the background disk at $a$, and  blackbody emission. The SED of the forming planet replaces this blackbody SED (in the absence of a gap). }
    \vspace{-15pt}
    \label{fig:SEDgrida}
\end{figure*}

\section{SED Results}\label{sec:SEDresults}

In this section, we present SEDs for a range of parameter values. We consider planet masses ranging from \qtyrange{0.33}{10}{M_J} and mass inflow rates ranging from \qtyrange{0.33}{10}{M_J\per\mega yr}. Over this range, the envelopes are optically thin and the parametric temperature approximation remains both accurate and efficient, {so we will use the parametric temperature approximation for the remainder of this work}. We specifically consider the effects of varying mass inflow rate, planet mass, viewing angle, opacity, and the semimajor axis of the orbit.  

\subsection{Mass Inflow Rate}

Fig. \ref{fig:SEDgridmdot} shows the spectral energy distribution $\nu L_\nu$ for forming planets with mass inflow rates that range from $\dot{M}=\qtyrange{0.33}{10}{M_J\per\mega yr}$. For larger mass inflow rates, the envelope radiation becomes dominant over the radiation of the disk and the planet. This trend results from the higher optical depth of the envelope, which causes significant attenuation of the planet and disk. At the low end of the range shown, few SED features distinguish between the different inflow distributions, i.e., the SEDs are largely independent of the geometric functions $f_i(\mu_0)$.  For larger infall rates, however, the envelope absorbs a larger fraction of the central source luminosity and its structure influences the SEDs to a greater extent. In particular, the SEDs are shifted to shorter wavelengths for polar inflow and shifted to longer wavelengths as the infall become concentrated in equatorial directions. These variations become evident for mass inflow rates of $\dot{M}\geq\qty{3}{M_J\per\mega yr}$. As expected, the total luminosity increases for higher mass inflow rates as more mass falls onto the central planet/disk system. 

H$\alpha$ observations of the protoplanets PDS 70 b/c suggest mass accretion rates in the range of $\dot{M}\simeq\qtyrange{0.01}{0.1}{M_J/\mega yr}$ \citep{Aoyama2019, Haffert2019, Thanathibodee2019, Hashimoto2020}. In Fig. \ref{fig:SEDgridlowMdot}, we show the spectral energy distributions for values of $\dot{M}$ consistent with these observations. As expected, the luminosity continues to decrease as the mass accretion rate is lowered. In addition, the disk and the envelope become dimmer in comparison to the planet as the total luminosity is reduced. Note that these SEDs may underestimate the planet's luminosity. In our derivation of the planet temperature and luminosity, we assume that the luminosity from mass accretion is significantly higher than the internal luminosity of the planet, but this assumption breaks down for sufficiently low mass accretion rates. For young Jupiter-mass objects, the internally-driven surface temperature is $T\sim\qty{750}{K}$ \citep{Chabrier2000}. As a result, the internal luminosity becomes important for mass accretion rates of $\dot{M}\lesssim\qty{0.03}{M_J/Myr}$ (the value where accretion produces similar planet temperatures -- see Eq. \ref{eq:Tp}). 

{Low mass accretion rates may also challenge the assumption that the circumplanetary disk is optically thick. At the outer edge of the circumplanetary disk, the optical depth of the disk (see Eq. 46 in \citealt{Taylor2024}) is approximately given by} 
\begin{equation}
    \tau_{d}=\num{300}\left(\frac{\dot{M}}{\qty{3}{M_J\per\mega yr}}\right)\left(\frac{\kappa_0}{\qty{10}{\centi\meter^2\per\gram}}\right)\,,
\end{equation}
{which applies for wavelengths near the peak of the SED. As a result, even for mass accretion rates as low as $\dot{M}=\qty{0.01}{M_J\per\mega yr}$, the disk remains optically thick ($\tau_d\simeq 10$). The disk begins to become optically thin (in the mid-infrared) for $\dot{M}\lesssim\qty{e-3}{M_J\per\mega yr}$, at which point the internal luminosity of the planet dominates over the accretion luminosity. In this regime, models of young giant planets that do not include  continuing accretion (e.g., \citealt{Marley2021}) are more applicable. Note that at long wavelengths, the opacity --- and hence the optical depth of the disk --- becomes small. Only a small fraction of the disk radiation is emitted at these wavelengths, however, so the SED is minimally affected.}  

{Although mass accretion rates as low as \qty{0.01}{M_J/\mega yr} are inferred for some protoplanets, we expect that mass accretion rates will typically be significantly larger. The runaway accretion phase of core accretion is expected to occur over only $\sim \qty{1}{\mega yr}$ or less, and giant planets formed through core accretion have masses of $M_p\sim \qty{1}{M_J}$. As a result, the average mass accretion rate during runaway accretion must be of order $\dot{M}\sim \qty{1}{M_J\per\mega yr}$, and much smaller mass accretion rates should be rare. The low accretion rates of these observed objects (most likely) reflect their advanced evolutionary state, since they exist in a deep gap within their parental protoplanetary disk. }

\subsection{Planet Mass}

Fig. \ref{fig:SEDgridm} shows the SEDs for a range of planet masses, from $M_p=\qty{0.33}{M_J}$ to $M=\qty{10}{M_J}$. We view these systems from along the planetary pole with all of the other system parameters set to their fiducial values. As expected, high-mass planets show a larger luminosity as the incoming material falls into a deeper potential well. Notably, the envelope luminosity does not generally increase at higher masses. The optical depth (equivalently column density) of the envelope is set primarily by the mass inflow rate, and the density decreases with increasing mass for constant ${\dot M}$. As a result, the SEDs for higher mass systems show relatively more emission at shorter wavelengths from the planet and its disk. Since the envelope produces most of the variations due to the inflow geometry, systems of lower mass have the most obvious geometric effects in their SEDs. 

\subsection{Viewing Angle}

Fig. \ref{fig:SEDgridangle} shows the SEDs for a range of viewing angles, from along the planetary pole ($\psi=0$) to near the equator ($\psi=3\pi/8$). Note that these systems are expected to be viewed along the planetary pole, as the background circumstellar disk will strongly attenuate the SED at oblique angles. Nevertheless, the behavior of the SEDs across viewing angle is as expected. In particular, for polar viewing angles the polar, quasipolar, and isotropic inflows produce SEDs with a broad, flat peak, while the quasiequatorial and equatorial SEDs exhibit a double-peaked distribution. For polar inflows, the envelope density is concentrated along the planetary pole relative to equatorial inflows. As a result, the envelope attenuates more luminosity when viewed from above than cases of equatorial inflow. In {most} cases, there are specific wavelengths where the disk {is brighter than} the envelope and the planet, providing a means of detecting the disk. However, a high optical depth can limit this wavelength range.

\subsection{Orbital Location}

Next we consider the effect of semimajor axis of the forming planet on its SED. In addition, we compare the SEDs of the forming system to the background SED of the protoplanetary disk (PPD). Since the forming planet system replaces the SED of the background disk, comparisons between these spectra are important to determine if the planetary system is detectable by contrast with the circumstellar disk. The temperature of the background disk is assumed to take the form 
\begin{equation}
    T_{\rm PPD}=T_{\rm X, PPD}\left(\frac{a}{R_{\rm X, PPD}}\right)^{-1/2}\,, \label{eq:TPPD}
\end{equation}
where $T_{\rm X, PPD}$ is the dust destruction temperature at \qty{1500}{K}. The radius $R_{\rm X, PPD}$ is the distance where this occurs, which we assume to be where the equilibrium temperature is $T_{\rm X, PPD}$. For a star of \qty{1.3}{L_\odot} (typical for a young solar-type star), this occurs at \qty{0.04}{au}. Note that circumstellar disks can have a range of temperature profiles, so this case is only an illustrative example. These results may be modified for known, specific disks. For this case, the SED of the background disk at semimajor axis $a$ will be
\begin{equation}
    L_{\rm PPD, \nu}=4\pi\,\pi R_H^2\,B_\nu(T_{\rm PPD}[a])\,,\label{eq:LPPD}
\end{equation}
where the factor of $4\pi$ accounts for the intensity-flux conversion and the factor of $\pi R_H^2$ is the area of the disk that is replaced by the planet system. Note that we are not accounting for a gap in the disk but rather are replacing the circumstellar disk within the Hill sphere with the SED of the forming planet. In this comparison, we assume that the planet SED is not attenuated by the background disk, leaving this matter for Sec. \ref{sec:backgrounddisk}. These curves are shown in Fig. \ref{fig:SEDgrida} along with corresponding SEDs of the planet/disk/envelope system. Note that since the background SEDs are independent of viewing angle or mass inflow rate, these curves are constant across all of the panels shown in Figs. \ref{fig:SEDgridmdot}--\ref{fig:SEDgridangle}. Since the luminosity scales as $L\propto R_H^2T^4$, $R_H\propto a$, and $T\propto a^{-1/2}$, the luminosity from the background disk is independent of $a$ for this model. 

Fig. \ref{fig:SEDgrida} shows that the forming planet system is generally dimmer than the background circumstellar disk. However, because the background disk is generally significantly cooler than the planet and its envelope, the system SED can be extracted from the background. Short-period planets will be difficult to detect, since the planet temperature and the background temperature of the circumstellar disk will be similar, and the background disk is bright. In contrast, forming planets with longer periods, $a\gtrsim\qty{1}{\astronomicalunit}$, will generally be easier to detect. 

\begin{figure}
    \centering
    \includegraphics{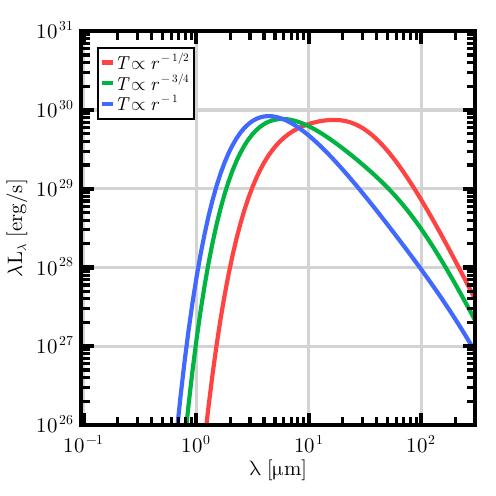}
    \vspace{-15pt}
    \caption{{The fiducial SED (viewed from above, with other parameters given by Table \ref{tab:canonvals}) with varying indices for the power-law temperature distribution of the disk. Results are shown for indices of $-1/2$ (red), $-3/4$ (green), and $-1$ (blue).}}
    \vspace{-15pt}
    \label{fig:tempscaling}
\end{figure}

\subsection{{Disk Temperature Distribution}}\label{subsec:disktempscale}

{Throughout this work, we have assumed that the disk temperature $T_d(r)\propto r^{-3/4}$, as is expected for an infinitesimally thin disk. However, previous work has considered alternate power-law indices ranging from $T\propto r^{-1/2}$ (characteristic of a flared disk) to $T\propto r^{-1}$ (suggested by some solar nebula models, \citealt{Lunine1982, Takata1996, Mosqueira2003a, Mosqueira2003b, Turner2014}). In Fig. \ref{fig:tempscaling}, we show the changes in the disk's radiative signatures for different temperature indices. Although the SEDs clearly vary due to different choices of the temperature distributions, these changes will have little effect on the radiative signatures of the entire system, since the planet and envelope generally dominate over the disk (and much of the disk emission is absorbed and reradiated by the envelope). Nonetheless, these models illustrate another possible source of variation in the SEDs of forming planets. }

\begin{figure}
    \centering
    \includegraphics{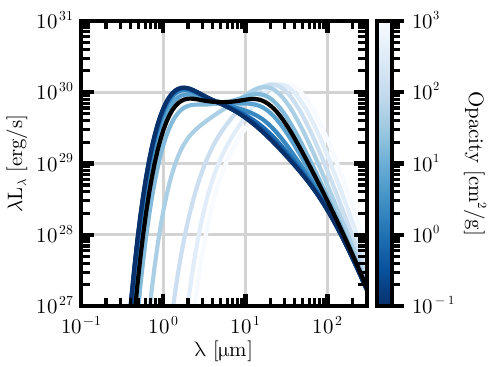}
    \vspace{-15pt}
    \caption{The fiducial SED (viewed from above, with other parameters given by Table \ref{tab:canonvals}) with varying values for the opacity coefficient. The black curve shows the SED in the case for $\kappa_0=\qty{10}{\centi\meter^2\per\gram}$.}
    \vspace{-15pt}
    \label{fig:opacitySED}
\end{figure}

\begin{figure}
    \centering
    \includegraphics{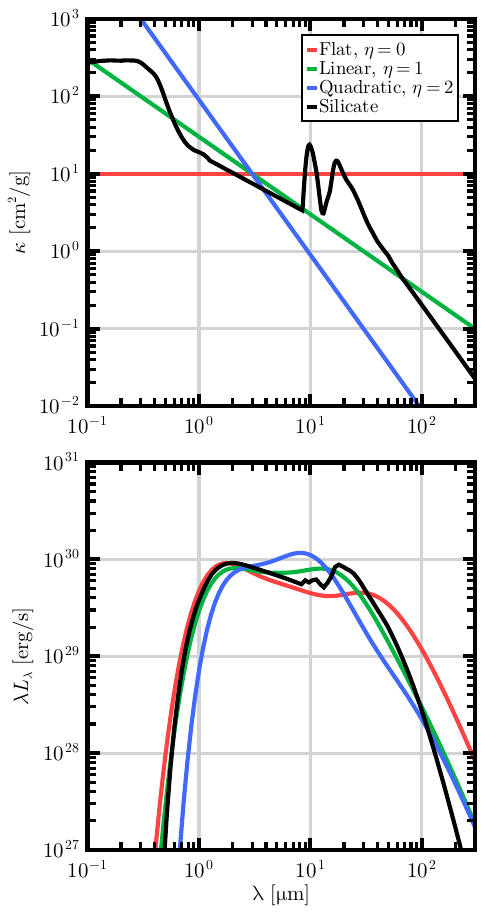}
    \vspace{-10pt}
    \caption{Comparison of opacities and the resulting SEDs. We show a constant opacity ($\eta=0$, red), a linear opacity ($\eta=1$, green), a quadratic opacity ($\eta=2$, blue), and the more-accurate (modified) silicate opacity (black). The top panel shows the opacity laws. The bottom panel shows SEDs calculated for these opacities. The system is viewed from the planetary pole and has fiducial parameter values (Table \ref{tab:canonvals}).}
    \vspace{-15pt}
    \label{fig:kappacomp}
\end{figure}

\subsection{Opacity Variations}\label{subsec:kappacomp}

\begin{figure*}
    \centering
    \includegraphics[width=\linewidth]{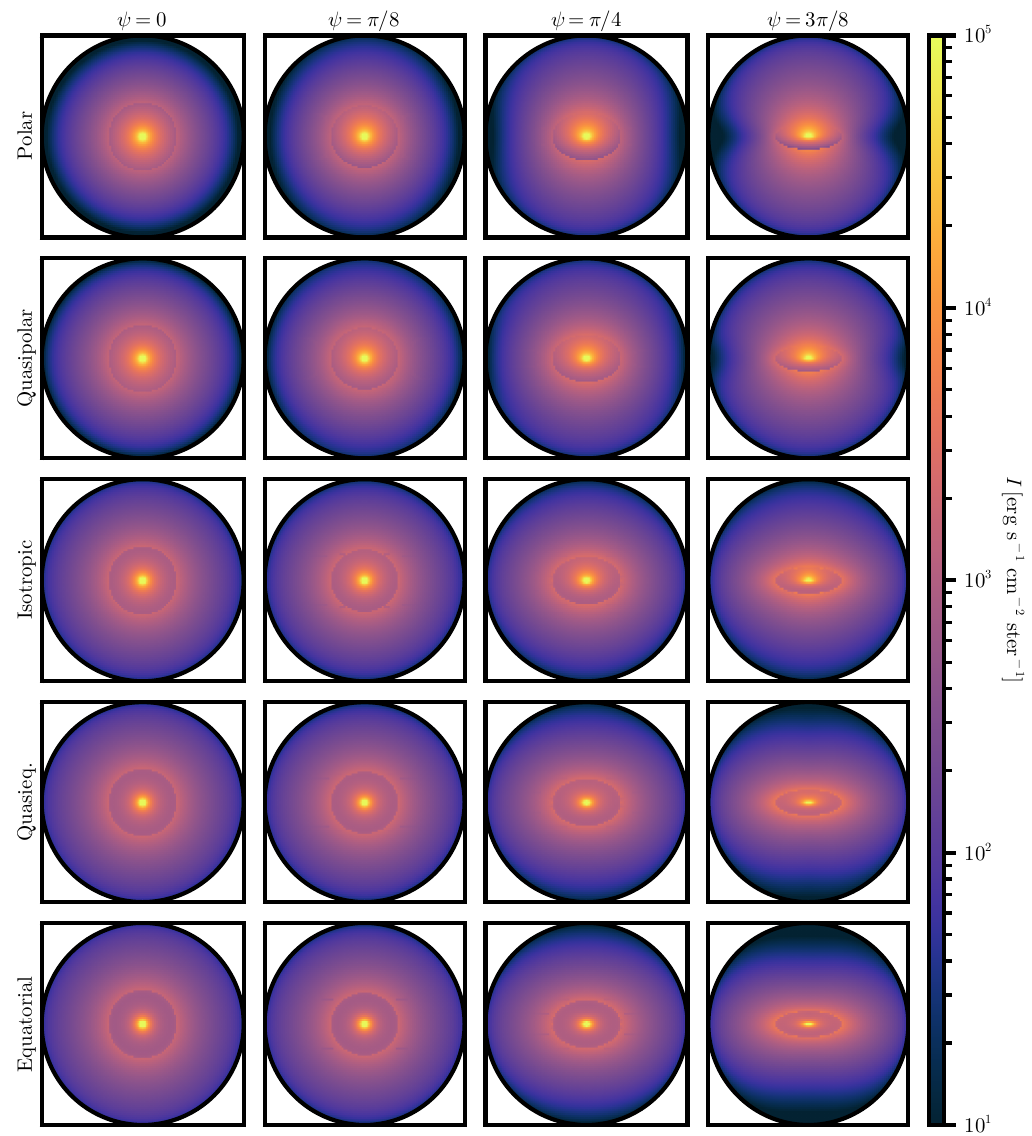}
    \vspace{-10pt}
    \caption{Synthetic images of the planet/disk/envelope system over a range of viewing angles and inflow functions, at a fixed mass flux of $\dot{M}=\qty{3}{M_J\per\mega yr}$. The domain extends to the Hill radius. All other parameters are given in Table~\ref{tab:canonvals}. The color shows the total intensity $I$ of the system along the ray at a given position. }
    \label{fig:imggridmass}
    \vspace{-15pt}
\end{figure*}

The opacity is one of the most important and uncertain parameters in determining the spectral signature of forming planets. While we have varied the mass inflow rate $\dot{M}$, the opacity constant $\kappa_0$ also affects the optical depth. More specifically, $\kappa_0\dot{M}$ sets the optical depth, although $\dot{M}$ also affects the luminosity and the column density. As a result, there is some degeneracy in the parameter space, and many of the variations that appear for increasing mass accretion rate (see Figs.~\ref{fig:SEDgridmdot} and \ref{fig:imggridmass}) will also appear for increasing opacity. For its fiducial value in this work, we have set $\kappa_0=\qty{10}{\centi\meter^2\per\gram}$, which is within the range advocated for circumstellar disks \citep{Semenov2003}. However, lower opacities are expected for circumplanetary disks as the planet formation process advances (e.g., \citealt{Weidenschilling1984}). As the opacity and optical depth decrease, the shape of the SED shifts toward shorter wavelengths, although the total luminosity is unmodified. 

Fig. \ref{fig:opacitySED} shows how the SED changes with variations in the opacity coefficient. Results are shown using fiducial values for the remaining parameters. These SEDs show that decreasing the opacity coefficient leads to a weakening in the envelope signature, allowing a larger fraction of radiation from the disk and planet to emerge. The total luminosity produced by the system remains constant. As a result, variations in the opacity modify the shape of the SED, with larger $\kappa_0$ resulting in redder SEDs. This type of SED shape modification is constrainted --- in the limit of zero opacity, the system SED is the sum of the planet and disk SED, without attenuation or re-emission from the envelope. For increasing opacity, the optical depth increases accordingly, and the envelope can become thick enough that the system falls outside the parameter space where the parametric approximation is valid. 

We also address the use of a linear opacity law in our calculations. Given the relatively limited temperature and wavelength range realized within the envelope, we have used an opacity law of the power-law form
\begin{equation}
    \kappa_\nu=\kappa_0\left(\frac{\nu}{\nu_0}\right)^\eta\,,
\end{equation}
where $0\leq\eta\leq2$. We generally set $\eta=1$, which is analytically tractable and captures the general behavior of the opacity over the wavelength range of interest. Fig. \ref{fig:kappacomp} compares the linear opacity law with a more-accurate opacity law (adopted from \texttt{RADMC-3D}) and shows the resulting SEDs. This more-accurate opacity law is a modification of the default silicate opacity in \texttt{RADMC-3D}, where we have included the effects of carbon grains by adding opacity in the wavelength region between \qty{1.25}{\micro\meter} and \qty{9}{\micro\meter} and used a gas-to-dust ratio of 100.

The resulting SEDs for the different opacity laws (Fig. \ref{fig:kappacomp}, panel 2), show small differences, primarily as a result of the \qty{10}{\micro\meter} silicate feature and the difference in the opacity at long wavelengths. These distinctions are modest, however, so that the linear opacity law provides a reasonable approximation. Moreover, although the primary distinction between these SEDs results from the \qty{10}{\micro\meter} silicate feature, the feature is not found in at least one known example of a circumplanetary disk \citep{Cugno2024}. 

We also compare SEDs for varying values of the opacity index $\eta$ in Fig. \ref{fig:kappacomp}. In these calculations, the index of the temperature distribution in Eq. \eqref{eq:Tesphere} is changed from $-2/5$ to $-2/(4+\eta)$. The corresponding changes in the SEDs are significant, but expected --- as the index increases, so does the total opacity (because we fix the overall value of $\kappa_\nu$ to be constant at \qty{3}{\micro\meter}). As a result, the attenuation from the envelope tends to contribute to a greater extent for larger values of the index $\eta$. Overall, the linear opacity law ($\eta=1$) provides the best match for the more-accurate opacity law, but the consequences of this assumption are shown clearly in Fig. \ref{fig:kappacomp}. Notice also that the differences in the opacity law itself (top panel of Fig. \ref{fig:kappacomp}) are more pronounced than those of the SEDs (bottom panel), as the latter result from integration over a range of $\kappa_\nu$.  

\begin{figure*}[t!]
    \centering
    \includegraphics[width=\linewidth]{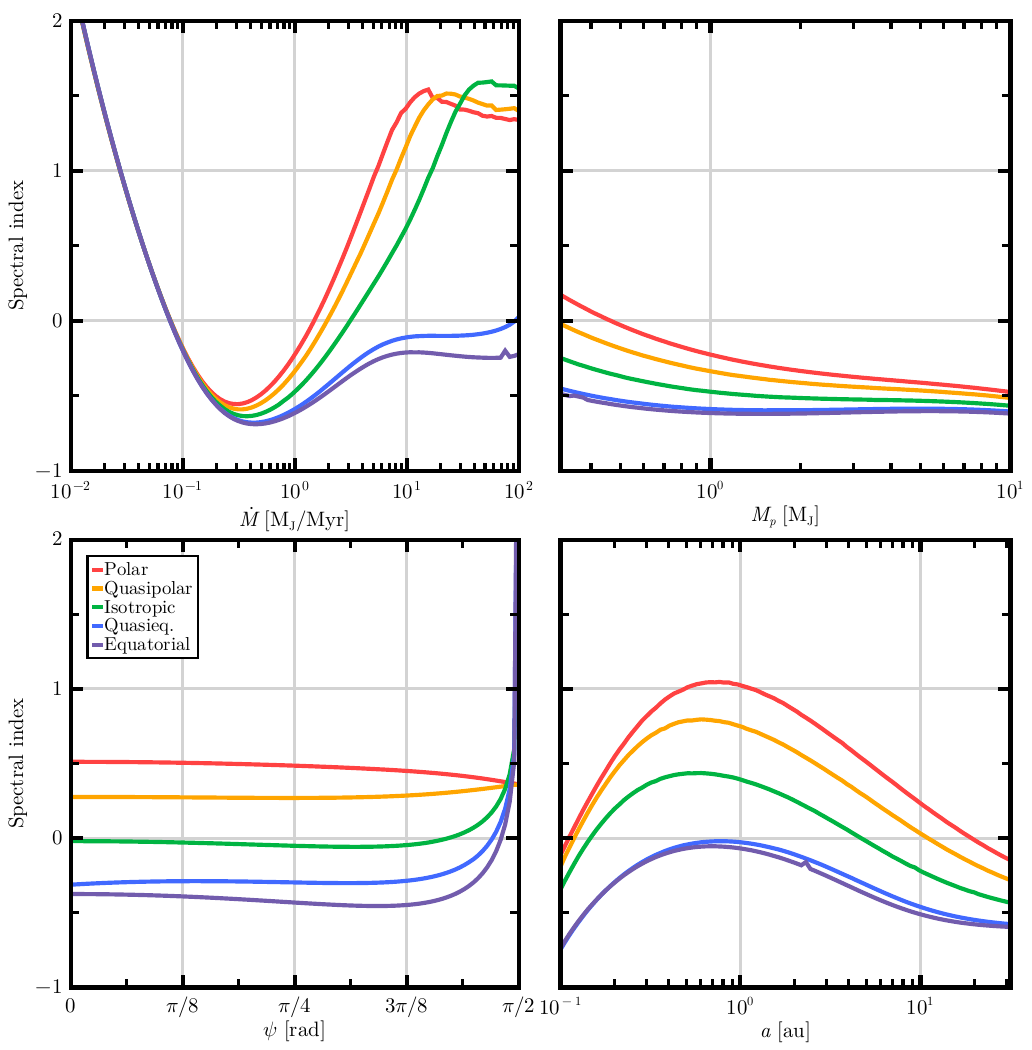}
    \vspace{-15pt}
    \caption{{The spectral indices of the SED (Eq. \ref{eq:specindex}) versus various model parameters. The colors show the different inflow geometries. We show the spectral index over the \qtyrange{2}{10}{\micro\meter} range. }}
    \vspace{-15pt}
    \label{fig:spectralindex}
\end{figure*}

\subsection{{The Infrared Spectral Index}}\label{subsec:spectralindex}

{We have presented SEDs over a large range of parameter space, which have varying slopes in IR bands. In this section, we propose the use of an IR spectral index for these disks, analogous to the index used for young stellar objects (\citealt{LadaWilk}; see Eq. (1) of \citealt{Adams1987}). Although there will necessarily be a large degeneracy in the values of this index across parameter space, observations of this index will likely be more accessible than full infrared spectra and may help to constrain the parameter space for individual systems.}

{At a particular wavelength $\lambda$ and corresponding frequency $\nu$, the spectral index $n$ is defined by}
\begin{equation}\label{eq:specindex}
    n\equiv\frac{\de\log (\lambda L_\lambda)}{\de\log(\lambda)}=-\frac{\de\log(\nu L_\nu)}{\de\log(\nu)}\,.
\end{equation}
{Note that we define the spectral index to be the negative of the traditional definition. We propose evaluation of the spectral index near the peak of the SEDs. Specifically, we define the IR spectral index $n$ to be the slope defined by Eq. \eqref{eq:specindex} evaluated over the wavelength range of \qtyrange{2}{10}{\micro\meter}. Fig. \ref{fig:spectralindex} shows this spectral index (calculated via a first-order finite difference method between the endpoints) versus a range of input parameters. Since we are using one number to characterize five parameters ($\dot{M}$, $M_p$, $\psi$, $a$, and inflow geometry), there is significant degeneracy.}

{Despite this degeneracy, Fig. \ref{fig:spectralindex} shows that this spectral index can help constrain the parameters of an accreting planetary system. The index characterizes the relative brightness of the planet and the envelope. If direct radiation from the planet dominates the system, then the spectral index traces the long wavelength regime of the planetary blackbody. As the envelope brightness grows, the spectral index increases. This behavior can be seen in the dependence of the index on the mass accretion rate at $\dot{M}\geq \qty{0.3}{M_J\per\mega yr}$, with the spectral index increasing as the envelope luminosity increases. A similar effect can be seen for $a\geq \qty{1}{au}$ --- at larger orbital distances, the Hill radius increases, the envelope cools, and the SEDs broaden. Increasing the mass of the planet increases the relative luminosity of the planet while decreasing the envelope column density, thereby lowering the spectral index. Variations in infall geometry are also captured by the spectral index, since the infall geometry primarily affects the density distribution and luminosity of the envelope. }

{At small mass accretion rates and orbital distances, however, the dependence of the spectral index on the mass accretion rate and orbital distance reverse. At sufficiently low mass accretion rates, the envelope emission becomes negligible, so that the planet contribution dominates the SED and the spectral index becomes that of the planetary blackbody. For lower mass accretion rates, the planetary surface becomes cooler and the blackbody peak shifts to longer wavelengths. As a result, the wavelength range over which the index is defined (\qtyrange{2}{10}{\micro\meter}) shifts from the long-wavelength end of the planet blackbody toward shorter wavelengths, thereby raising the index (see Fig. \ref{fig:spectralindex}). A similar argument explains the decrease in the spectral index at small semimajor axes. As the orbital distance decreases, the envelope tends to dominate the system's emission. Simultaneously, however, the envelope temperature increases, eventually shifting the wavelength range of the index from the long-wavelength side of the SED to the short-wavelength side. }


\begin{figure*}
    \centering
    \includegraphics[width=\linewidth]{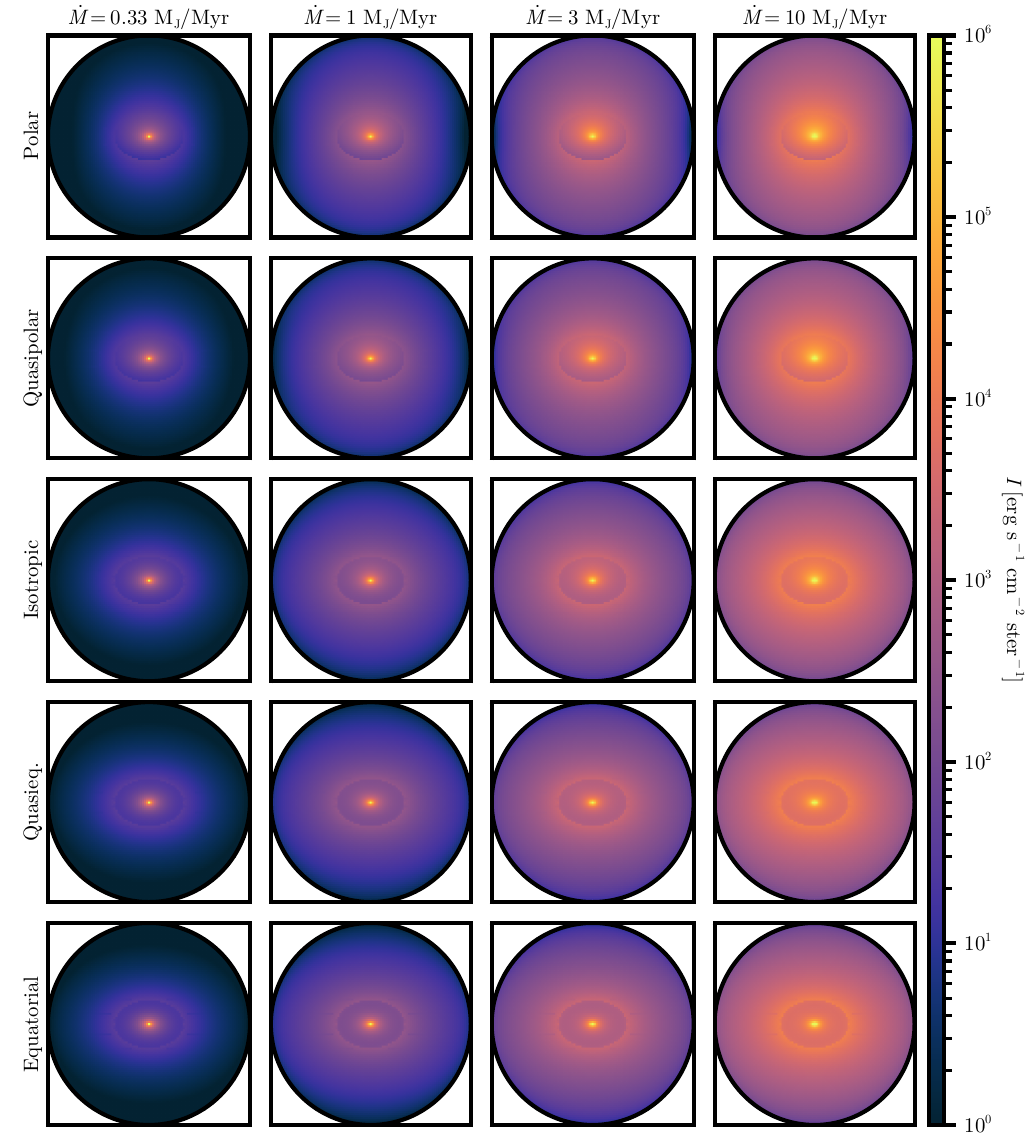}
    \vspace{-10pt}
    \caption{Synthetic images of the planet/disk/envelope system over a range of mass inflow rates and inflow functions, at a fixed viewing angle of $\psi=\pi/4$. The domain extends to the Hill radius. All other parameters are given in Table~\ref{tab:canonvals}. The color shows the total intensity $I$ of the system along the ray at a given position.  }
    \vspace{-15pt}
    \label{fig:imggridangle}
\end{figure*}

\section{Synthetic Images}\label{sec:images}

In this section, we construct synthetic images of the planet/disk/envelope system. These images are constructed to be the intensity at a detector plane at infinite distance, as defined in Sec.~\ref{sec:SEDs}. We here present the total intensity of radiation passing through each point on the detector plane, although modifications can be made for specific wavelength bands. For convenience, we define a circular coordinate system on the detector plane --- $b$ is the distance from the center of the plane and $\varphi$ is the azimuthal angle, which is set to be $0$ when the point is directed towards the planetary system's pole. For each point on the image defined by ($b,\varphi$), we trace a ray through the planet/disk/envelope system. The total intensity along this ray can be separated into components from the planet, disk, and envelope, so that the intensity is given by
\begin{equation}
\begin{split}
    I &= \int\displaylimits_0^\infty\!\!\de\nu \left(I_{p,\nu}+I_{d,\nu}+I_{e,\nu}\right)\\
      & = I_p + I_d + I_e \,. \label{eq:Itotal}
\end{split}
\end{equation}
These components are evaluated in turn.

\subsection{Planet Intensity}

Any ray that strikes the planet carries an intensity of $I_{p,\nu}=B_\nu(T_p)\,\exp(-\kappa_\nu N_{\rm col})$, where $N_{\rm col}$ is the column density between the Hill sphere and the planetary surface. Integrating this over frequency, we can determine that 
\begin{equation}
    I_p=\frac{1}{\pi}\sigma T_p^4\,\alpha_P(T_p, N_{\rm col})\,. \label{eq:Ip}
\end{equation}
In Eq. \eqref{eq:Ip}, $\alpha_P$ is the effective Planck-weighted attenuation coefficient, which is derived in Appendix \ref{subsec:alphaderiv}. This intensity is only added to rays that strike the planetary surface. Since the traced rays are parallel to the detector plane, the ray will only strike the planet if $b\leq R_p$. In this case, the planet's intensity is given by Eq. \eqref{eq:Ip}; otherwise $I_p=0$. The column density is assumed to be constant for all points on the planet's surface and is written as 
\begin{equation}
    N_{\rm col}=\int\displaylimits_{R_p}^{R_H}\!\!\rho(r,\,\cos\psi)\,\de r\,,
\end{equation}
where $\rho$ is given by Eq. \eqref{eq:envrho}.

\subsection{Disk Intensity}

Like the planet, the disk is assumed to be a blackbody emitter, albeit at a variable temperature. If the ray crosses through the equatorial plane at a radial distance $r_0$, the intensity from this point is given by 
\begin{equation}
    I_d=\frac{1}{\pi}\sigma T_d^4[r_0]\,\alpha_P\!\left[T_d(r_0), N_{\rm col, d}\right]\,, \label{eq:Id}
\end{equation}
where the disk temperature is given by Eq. \eqref{eq:Td} and $N_{\rm col, d}$ is the column density between the viewing point and the disk. We then calculate $r_0$ and $N_{\rm col, d}$ for a given detector plane point $(b, \varphi$). Using the Cartesian formulation from Sec. \ref{sec:SEDs}, we find that 
\begin{equation}
    r_0=b\sqrt{1+\cos^2\varphi\tan^2\psi}\,.\label{eq:r0def}
\end{equation}
Next consider the column density. Eqs. \eqref{eq:diskrays} and \eqref{eq:dNcol} specify the column density between the disk and the Hill sphere for a given point on the disk surface. We have already determined that our intersection point lies at $r'=r_0$, so we now find the cosine of azimuthal angle at the disk intersection point. A similar argument indicates that 
\begin{equation}
    \cos\phi=-\cos\varphi\sqrt{\cos^2\psi+\sin^2\psi\cos^2\varphi}\,. \label{eq:diskphicalc}
\end{equation}
By combining Eqs. \eqref{eq:r0def} and \eqref{eq:diskphicalc} with Eqs. \eqref{eq:Id} and \eqref{eq:Td}, we can calculate the intensity due to the disk at every point. Since the disk has a finite extent, if $R_X\leq r_0\leq R_C$, then $I_d$ is given by Eq. \eqref{eq:Id}; otherwise $I_d=0$.

\subsection{Envelope Intensity}

Consider a ray of differential length $\de s$ passing through a point in the envelope with temperature $T_e$. The intensity added to this ray by the envelope material is given by 
\begin{equation}
\begin{split}
    \de I_e(r, \mu) &= \int\displaylimits_0^\infty\!\!\de\nu\,\rho(r, \mu)\,\kappa_\nu B_\nu[T_e(r,\mu)]\,\de s\\
        &= \frac{1}{\pi}\rho(r, \mu)\,\kappa_P\sigma T_e^4[r,\mu]\,\de s\,. \label{eq:Iepoint}
\end{split}
\end{equation}
As a result, the envelope intensity at each point in the detector plane must be determined by an integral along the corresponding ray. Each ray will describe a chord through the sphere, with the closest approach to the center at a distance $b$. By defining a parameter $s$ along the chord, the total intensity at a point ($b,\varphi$) is given by 
\begin{equation}
    I_e(b,\,\varphi)=\frac{1}{\pi}\sigma\!\!\!\int\displaylimits_{s_{\rm min}}^{s_{\rm max}}\!\!\!\de s\,\rho\!\left[r(s),\mu(s)\right]\kappa_P \,T_e^4\!\left[r(s),\mu(s)\right]\,.\label{eq:Ie}
\end{equation}

In Eq. \eqref{eq:Ie}, the integral is bounded by the edges of the chord. Since the upper end of the chord is always at the Hill sphere, the maximum value of $s$ is
\begin{equation}\label{eq:smax}
    s_{\rm max}=\sqrt{R_H^2-b^2}\,.
\end{equation}
The minimum value is more complicated. If the chord passes through the entire sphere, then the symmetry of the chord means that $s_{\rm min}=-s_{\rm max}$. However, if the ray strikes the planet or the disk, then the ray is truncated at that point, since these components are opticaly thick. If the ray strikes the planet (i.e., if $b\leq R_p$) then $s_{\rm min}=[R_p^2-b^2]^{1/2}$. Alternatively, the ray will strike the disk if $R_X\leq r_0\leq R_C$, where $r_0$ is the distance from the origin when the chord crosses the equatorial plane and is given by Eq. \eqref{eq:r0def}. In this case, the minimum is $s_{\rm min}=b\cos\varphi\tan\psi$. Therefore, the lower bound of the integral in Eq. \eqref{eq:Ie} is given by
\begin{equation}\label{eq:smin}
    s_{\rm min}=\begin{dcases}
        \phantom{-}\sqrt{R_p^2-b^2} & b\leq R_p\,;\\
        \phantom{-}b\cos\varphi\tan\psi & R_X\leq r_0\leq R_C\,;\\
        -\sqrt{R_H^2-b^2} & \text{otherwise.}
    \end{dcases}
\end{equation}

In order to evaluate the envelope density and temperature at a given point along the ray, the radius $r$ and the cosine $\mu$ of the polar angle must be expressed as a function of the variable $s$. We find that 
\begin{subequations}\label{eq:sphereraypts}
\begin{align}
    r&=\sqrt{b^2+s^2}\,;\\
    \mu&=\frac{s+b\sin\psi}{r}\cos\varphi\,.
\end{align}
\end{subequations}
Eqs. \eqref{eq:smax}, \eqref{eq:smin}, and \eqref{eq:sphereraypts} can then be combined with Eq. \eqref{eq:Ie} to find the intensity due to the envelope at a point on the detector. 

Eq.~\eqref{eq:Itotal} can now be used to compute the intensity for each point of the detector plane from the corresponding normal ray. In Figs.~\ref{fig:imggridangle} and \ref{fig:imggridmass}, we show the resulting images over a range of parameter-space values for a $101\times101$ grid on the image plane. 

These images show a general increase in luminosity for higher mass inflow rate, as expected. For lower mass inflow rates, the images show few distinctions, since the luminosity and the column density are low enough that the emission is largely spatially diffuse. At high mass inflow rates, there are some observational distinctions between the different inflow geometries. The concentration of incoming material along the pole causes greater absorption of the planetary luminosity in this region, so the central source becomes larger and more diffuse. In contrast, equatorial inflows show the central source as a point, reflecting the small size of the planet. 

It is also notable that in all cases, the disk appears dark at the outer edge, indicating relatively low temperatures. In general, the envelope appears slightly brighter just outside the outer edge of the disk. This feature is due to the structure of the infalling envelope, which has a pile-up of material and hence a local density maximum near the centrifugal radius (see also \citealt{Ulrich1976,Cassen1981}). The increased density of the envelope results in the bright band shown in the emission maps. In addition, the ray along the line of sight is stopped at the disk, so that nearly twice as much radiation is visible just outside the disk than just inside of it. The presence of a disk is therefore visually clear in the images, for all values of the mass inflow rate and column density. 

\begin{figure}
    \centering
    \includegraphics{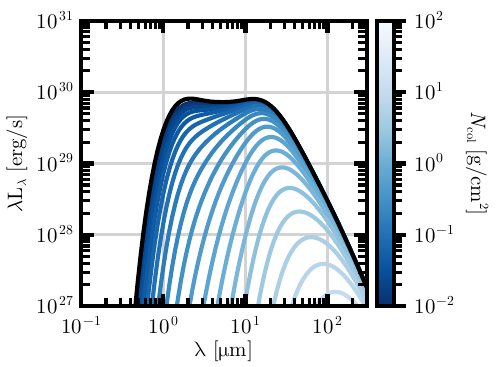}
    \vspace{-10pt}
    \caption{The fiducial SED (viewed from above, with other parameters given by Table \ref{tab:canonvals}) with varying levels of attenuation due to the circumstellar disk. The black curve shows the SED in the absence of attenuation.}
    \vspace{-15pt}
    \label{fig:PPDatt}
\end{figure}

\section{Background Protoplanetary Disk}\label{sec:backgrounddisk}

This section considers the effects of the background circumstellar disk on the SEDs of the forming planet/disk/envelope system. This background disk will not only feed the forming planet by funneling material into the Hill sphere, but will also attenuate the outgoing radiation of the system.

First, we calculate the effect that the background circumstellar disk would have on the SEDs of the system for the case where the planet remains fully embedded. For a minimum-mass solar nebula (MMSN), the surface density \citep{Hayashi1981} is given by 
\begin{equation}
    \Sigma=1.752\times 10^3\left(\frac{a}{\rm au}\right)^{-3/2}\,\text{g cm}^{-2}\,.\label{eq:MMSN}
\end{equation}
At a distance of $a=5$ au from the central star, the system is viewed through a column density of approximately $N_{\rm col}\simeq\Sigma/2=\qty{157}{\gram\,\centi\meter^{-2}}$. Fig. \ref{fig:PPDatt} shows the system SED (for the fiducial parameters in Table \ref{tab:canonvals}) including further attenuation by a factor of $\exp(-\kappa_\nu N_{\rm col})$, which accounts for the surrounding disk. This figure shows that if the full column density of the MMSN is present, the SED of the forming planet would be so strongly attenuated that it would not be detectable. As a result, the surface density (of the background circumstellar disk) must be reduced in order to render the planet observable.

Giant planets are likely to clear gaps in the circumstellar disk during the late stages of their formation. The resulting density suppression in the gap can be written in the form \citep{Kanagawa2015, Paardekooper2023}
\begin{equation}
    \frac{\Sigma_{\rm gap}}{\Sigma_0}=\frac{1}{1+q^2/(29h^5\alpha)}\,,\label{eq:suppression}
\end{equation}
where $q=M/M_\star$, $h=H/a$, $H$ is the scale height of the disk, and $\alpha$ is the usual viscosity parameter (starting with \citealt{Shakura1973}). Since $q\sim10^{-3}$ and $h\sim1/20$, the expression for gap clearing \eqref{eq:suppression} takes the approximate form
\begin{equation}
    \frac{\Sigma_{\rm gap}}{\Sigma_0}\simeq10\alpha\,,
\end{equation}
which is valid in the (expected) limit where the viscosity parameter $\alpha$ is small.  

As benchmark for comparison, we consider the planet to be potentially observational when the surface density $\Sigma$ of the background circumstellar disk is less than the column density of the planetary envelope. For isotropic inflow onto the planet and its disk, the spherically averaged column density of the envelope is given by 
\begin{equation}
\begin{split}
    N_{\rm col}=0.055&\left(\frac{\dot{M}}{\unit{M_J\per\mega yr}}\right)\left(\frac{M_\star}{\unit{M_\odot}}\right)^{1/6}\\
    \times&\left(\frac{M}{\unit{M_J}}\right)^{-2/3}\left(\frac{a}{\unit{\astronomicalunit}}\right)^{-1/2}\unit{\gram\,\centi\meter^{-2}}\,.\label{eq:Ncol}
\end{split}
\end{equation}
Combining Eq. \eqref{eq:Ncol} with Eq. \eqref{eq:MMSN}, we find that
\begin{equation}
\begin{split}
    \frac{N_{\rm col}}{\Sigma}=0.063&\left(\frac{\alpha}{\num{e-4}}\right)^{-1}\left(\frac{\dot{M}}{\unit{M_J\per\mega yr}}\right)\\
    \times&\left(\frac{M}{\unit{M_J}}\right)^{-2/3}\left(\frac{a}{\unit{\astronomicalunit}}\right)\,.   
\end{split}
\end{equation}
In order for the optical depth of the envelope to be larger than the optical depth of the background disk, we require that
\begin{equation}
    \dot{M}\geq\num{6.45e4}\alpha\,\unit{M_J\per\mega yr}\,.
\end{equation}
Since we expect ${\dot M}\sim\qtyrange{1}{10}{M_J\per\mega yr}$, background disks with $\alpha\sim\num{e-4}$ will produce cleared gaps with surface densities comparable to the column density of the envelopes of forming planets. 

In addition, the viscous migration timescale is given by 
\begin{equation}
    \tau_{\rm acc}=\frac{a^2}{\nu}=\frac{a^2}{\alpha H c_s}\simeq\frac{\qty{1000}{yr}}{\alpha}\,.
\end{equation}
In order for these planets to form readily, the accretion time scale must be shorter than (or perhaps comparable to) their migration timescale, so that $\tau_{\rm acc}\geq M/\dot{M}$. This constraint can be written as
\begin{equation}
    \dot{M}\gtrsim\num{e3}\alpha\,\unit{M_J\per\mega yr}\,.
\end{equation}
As a result, the restriction for sufficient gap opening (and surface density reduction) is somewhat more stringent than the restriction for planetary survival. However, large values of $\dot{M}$ will lead to an optically thick planetary envelope, obscuring the planet disk. Smaller viscosities are therefore a preferable path to ensure that the circumstellar disk is optically thin relative to the protoplanetary system.

Planets on longer-period orbits and in systems with smaller effective viscosities will be less attenuated by their background circumstellar disks. In addition, given that a gap is required for the planet to be observable, an observed planet will necessarily be large enough to clear a gap. Since the mass accretion rate of the forming planet increases with mass, ${\dot M}$  will also be large, and the formation timescale will be relatively short. These results imply that relatively few planets will be observable during the formation process, or equivalently, forming planets will be observable for limited spans of time, which may be relevant for population calculations.

\section{Conclusion}\label{sec:disc}

\subsection{Summary}

In this paper, we have calculated spectral energy distributions (SEDs) and synthetic images for giant planets in the late stages of formation. In this phase, the planet is accompanied by a circumplanetary disk and is embedded within a circumplanetary envelope, which is itself surrounded by the background circumstellar disk. We build on previous work that constructed {a} semianalytic model for the planet/disk/envelope system over a range of boundary conditions that set the distribution of infalling material at the Hill radius \citep{Adams2022,Taylor2024}. {We have improved upon this previous work by (i) relaxing the assumptions that were previously used (spherically symmetric envelope density and a point-source disk) and (ii) rigorously testing the remaining assumptions (spherically symmetric envelope temperature distribution).}  Using this model, we have obtained radiative signatures of forming planets that may be detectable in the near future. Our specific results are outlined below.  

Although the allowed parameter space for forming planets is vast and leads to a wide range of possible radiative signatures, the basics properties of the SEDs can be summarized as follows. For planetary masses of order $M\sim\qty{1}{M_J}$ and expected infall rates ${\dot M}\sim\qtyrange{1}{10}{M_J\per\mega yr}$, most of the radiation is emitted at near to mid-infrared wavelengths $\lambda=\qtyrange{1}{100}{\micro\meter}$ (see Fig. \ref{fig:SEDgridmdot}). The intrinsic brightness of these forming objects are characterized by $\lambda L_\lambda \sim\qty{e30}{erg\per\second}\simeq \qty{e-3}{L_\odot}$. These power outputs are about 4 orders of magnitude fainter than the young star/disk systems that form the planets. While star/disk systems have been observed for decades (e.g., \citealt{LadaWilk}), planet/disk systems are just now becoming observable (e.g., \citealt{Benisty2021,Cugno2024}). Notably, these luminosities are sufficiently bright to be detectable at distances of \qty{150}{pc}, which encompasses many star-forming regions where these systems may be found (Taurus, Orion, Lupus, etc). 

Variations in the infall geometries, as specified through the function $f_i(\mu_0)$ defined in Eq. \eqref{eq:asyminfunc}, lead to qualitative and quantitative differences in the density structure of the envelopes. These variations affect both the SEDs and the synthetic images, especially for high luminosities and optical depths (see Figs. \ref{fig:SEDgridmdot}, \ref{fig:SEDgridm}, \ref{fig:SEDgridangle}, and \ref{fig:imggridmass}). For systems viewed from the polar directions, which are most easily detectable through the background circumstellar disk, the SEDs become bluer (redder) as the infall becomes increasingly concentrated toward the equatorial (polar) directions (Fig. \ref{fig:SEDgridangle}). 

The mass accretion rate ${\dot M}$ affects the SEDs in two significant ways. As ${\dot M}$ increases, the total luminosity of the system increases (see Fig. \ref{fig:SEDgridmdot}). In tandem, the optical depth of the envelope increases, which can cause the envelope to dominate the spectral signatures (depending on the mass inflow rate) and leads to redder SEDs. Increasing the mass inflow rate and/or the optical depth also causes the SEDs to vary more substantially across infall geometries. 

As the mass increases, the system luminosity increases due to the greater depth of the gravitational potential well. As a corollary, the increased luminosity from the planet and the disk requires increased temperatures and hence bluer SEDs (see Fig. \ref{fig:SEDgridm}). The envelope optical depth does not increase, so emission from the envelope is relatively unmodified. As a result, planets with the smallest masses show the largest qualitative differences in their SEDs due to variations in the infall geometry, as the infall geometry primarily affects the envelope. As the planet mass increases, the size of the Hill sphere increases accordingly, leading to a wider range of envelope temperature and a corresponding broadening of the SED. 

Varying the viewing angle does not significantly modify the SEDs (see Fig. \ref{fig:SEDgridangle}). The primary effect is that the direct emission from the circumplanetary disk becomes weaker with increasing polar angle, i.e., the disk contribution to the SED is reduced by a factor of $\cos\psi$. This variation has relatively little effect on the overall result, since the planet and envelope generally dominate the SEDs at these optical depths. Note that a wide range of viewing angles is not expected, since large polar angles imply sight lines that intersect the background circumstellar disk and become heavily attenuated. 

The orbital location of the planet is a more significant factor. As the planet orbits farther from its parental star, the Hill radius grows larger. This increase causes the disk and the envelope contributions to the SEDs to broaden, since a greater range of temperatures are sampled in this expanded configuration (see Fig. \ref{fig:SEDgrida}). For fixed mass inflow rate, the larger Hill radius leads to a lower total column density and allows for more of the planet and disk radiation to escape the system. This effect is similar to increasing the planetary mass, which enlarges the Hill sphere in analogous fashion. Distinguishing the planetary system from the background disk requires a relatively high system luminosity and/or additional spectral information. In general, radiation from the forming planet is bluer than that of the background disk, although this ordering is not always the case for short-period planets. High mass inflow rates and/or large planet masses are necessary for the system to be brighter than the background disk, and this consideration further restricts the regime over which forming planets are observable. 

The total optical depth of the infalling envelope plays a critical role in the potential observability of forming planets. While the envelope optical depth has no effect on the total system luminosity, it significantly reshapes the emergent spectral energy distribution. The different possible infall geometries (from polar to equatorial flow) result in varying optical depths, which thus provide an important signpost of these boundary conditions. Note that we are working within the (expected) regime where the viscosity of the circumplanetary disk can efficiently funnel most of the incoming material onto the planet. In this limit, the location where the material first lands on the disk is effectively erased, and the disk surface density takes nearly the same functional form for all infall geometries. 

{We propose characterizing the SEDs using a spectral index defined in the near-IR over the wavelength range \qtyrange{2}{10}{\micro\meter}. This index captures the relative contributions of the planet and the envelope to the SEDs. It can thus constrain the model parameters for accreting planets and provide insight into the structure of the planet/disk/envelope system (see Fig. \ref{fig:spectralindex}). The spectral index requires less data than a full SED and is thus more accessible. This index may provide an important tool for characterizing young and forming giant planets in the future.  }

Most of the results presented in this paper use a parametric approximation, where the temperature distribution of the envelope is spherically symmetric and the density distribution has its full aspherical form. For the parameter space under consideration, this approximation is both accurate and efficient, allowing for rapid evaluation of the SED and future model fitting. These assumptions begin to break down at sufficiently high optical depths, which manifest at high infall rates $\dot{M}\gtrsim\qty{30}{M_J\per\mega yr}$ (see Figs. \ref{fig:tempgrids} and \ref{fig:SEDMCcomp}). Above this threshold, the envelope becomes optically thick to its internal radiation field and numerical radiative transfer calculations (here we use \texttt{RADMC-3D}) provide greater accuracy. At these high optical depths, however, the system develops an effective photosphere near the Hill radius, and the intrinsic radiation from the planet and disk become highly obscured. In the optically thick regime, few details of the planet and disk properties remain in the SED, which approaches a blackbody form (see Fig. \ref{fig:SEDMCcomp}). 

The SEDs of forming planets are greatly affected by the degree of gap clearing in the background circumstellar disk. For disk properties comparable to the minimum-mass solar nebula, and in the absence of a gap, the background circumstellar disk would heavily attenuate the system's radiative signatures (see Fig. \ref{fig:PPDatt}). In this limit, the forming planet would be nearly unobservable. In order for forming giant planets to be detected, the system must clear a gap so that the optical depth of the background circumstellar disk is smaller than (or perhaps comparable to) the optical depth of the infalling envelope surrounding the planet. These conditions are expected in the later stages of the planet formation process. 

\subsection{Discussion}

We have found that the different infall geometries and/or envelope density distributions lead to qualitative and quantitative differences in the SEDs and synthetic images, especially for high luminosities and optical depths. While images of the system more readily show the presence of the disk, observations with sufficient angular resolution to distinguish these features are not possible in the near future. Instead, upcoming spectral observations are more likely to provide constraints on system properties, including the presence of a disk and the density distributions of the envelopes.  

The results of this paper indicate certain strategies for future observations. The circumplanetary disk will generally provide the dominant radiation signal in the near infrared ($\lambda\sim\qty{5}{\micro\meter}$). As a result, observations hunting for such disks should focus on near-infrared wavelength bands. Note that this work does not account for line emission, which may provide a more distinct signature of the circumplanetary disk, and should be pursued in future work. Notice also that the disk emission is more visible as the optical depth of the infalling envelope decreases, which occurs as the planet mass grows (for constant ${\dot M}$). This trend, in combination with gap clearing in the background circumstellar disk, implies that the circumplanetary disk will be most visible in the extreme late stages of the planet forming process. {Observations of the IR spectral index discussed in Sec. \ref{subsec:spectralindex} will also help to constrain the properties and structure of the protoplanetary system.}

The opacity represents a significant source of uncertainty in determining the spectral energy distributions and radiation maps. While the interstellar dust opacity has been well studied (from \citealt{Draine1984} onward), the dust present in the envelopes of forming planets is expected to be highly evolved from its interstellar origins. The dust settles to the midplane of the circumstellar disk, and coagulates into larger bodies, thereby reducing the effective opacity (e.g., \citealt{Weidenschilling1984}). In addition, if the dust grains grow large enough, they become decoupled from the gas and are subject to migration and other effects. All of these potential changes to the opacity result in corresponding changes in the SEDs, which we do not address. On the other hand, the range of temperatures sampled by these systems ranges from $T\sim1000$ K at the planetary surface down to $T\sim\qty{100}{K}$ at the Hill sphere. As a result, the relevant wavelength range for the opacity is only about one order of magnitude, so that a simplified treatment is justified. We have (primarily) used a power-law prescription for the opacity and showed how the resulting SEDs change with the opacity coefficient (Fig. \ref{fig:opacitySED}), opacity index, and the presence of a silicate feature (Fig. \ref{fig:kappacomp}). As more observations become available for comparison, this treatment can be expanded to include more detailed opacity laws, although the possible parameter space is large. 

In our calculations, we have also assumed azimuthal symmetry. While this is generally justified due to the short orbital periods in the planetary environment, streaming infall and modifications of the envelope density structure due to hydrodynamic effects may significantly modify the radiative signatures of these systems. 

These models can be used to provide insight into the radiative signatures of forming planets as they evolve. Since the SED is a nearly-instantaneous tracer of the system's structure, evolutionary sequences can be constructed using these results. As the planet evolves, the mass inflow rate, planet mass, opacity, semimajor axis, and infall geometry may change. In addition, forming planets are expected to induce structure in the background protoplanetary disk through their dynamical influence. Gap structures are particularly important, as the planet depletes material from the local disk. Once a gap has been opened, it is likely that equatorial inflows will be preferred --- the infalling material will have to cross the gap and flow through the equator, rather than leaping out of the disk plane and entering through the planetary pole. As a result, a protoplanet will likely change its inflow distribution over its formation lifetime, leading to a more complex disk structure and evolutionary history. Due to its efficiency, the approach developed in this paper is a useful tool for exploring the radiative signatures of these planets over their evolutionary history. 

In the future, several improvements should be made to these SED calculations. For much of this work, we have assumed a power-law absorption and emission opacity, lacking any spectral lines or additional features from molecular composition, particle geometry, or accretion shocks. Such features may provide more precise details on the parameters of the system and will refine the predicted signatures presented here. This paper has also generally neglected magnetic fields, only invoking them to define the inner edge of the disk. In addition, however, stronger magnetic fields result in an increased cross section for the planet to directly capture incoming material \citep{Adams2022}, leading to larger planetary luminosity. Magnetic fields can also lead to magnetic braking and angular momentum transfer, which in turn affects both disk and envelope structure. The infall solutions that determine the envelope density distribution can also be generalized to include pressure effects, more complicated boundary conditions, and the loss of azimuthal symmetry. In any case, this paper represents a starting step toward understanding the radiation signatures of planet formation. As observations of forming planets (along with their disks and envelopes) are carried out, SED models will require ever-increasing complexity to ascertain system properties. 

\section*{Acknowledgements}

{We thank the two anonymous reviewers for their helpful comments.} We thank Konstantin Batygin, Nuria Calvet, Gabriele Cugno, Lee Hartmann, Thomas Kennedy, Kaitlin Kratter, and Michael Meyer for helpful conversations. A.G.T. acknowledges support from the Fannie and John Hertz Foundation and the University of Michigan's Rackham Merit Fellowship Program. This research was supported in part through computational resources and services provided by Advanced Research Computing at the University of Michigan, Ann Arbor. This paper made use of the Julia programming language \citep{Julia}, the plotting package Makie \citep{Makie}, and Wolfram Mathematica \citep{Mathematica}.

\bibliographystyle{aasjournal}
\bibliography{main.bib}

\begin{thebibliography}{}
\expandafter\ifx\csname natexlab\endcsname\relax\def\natexlab#1{#1}\fi
\providecommand{\url}[1]{\href{#1}{#1}}
\providecommand{\dodoi}[1]{doi:~\href{http://doi.org/#1}{\nolinkurl{#1}}}
\providecommand{\doeprint}[1]{\href{http://ascl.net/#1}{\nolinkurl{http://ascl.net/#1}}}
\providecommand{\doarXiv}[1]{\href{https://arxiv.org/abs/#1}{\nolinkurl{https://arxiv.org/abs/#1}}}

\bibitem[{{Abramowitz} \& {Stegun}(1972)}]{AbramStegun1972}
{Abramowitz}, M., \& {Stegun}, I.~A. 1972, {Handbook of Mathematical Functions} (Dover Books)

\bibitem[{{Adams} \& {Batygin}(2022)}]{Adams2022}
{Adams}, F.~C., \& {Batygin}, K. 2022, \apj, 934, 111, \dodoi{10.3847/1538-4357/ac7a3e}

\bibitem[{{Adams} {et~al.}(1987){Adams}, {Lada}, \& {Shu}}]{Adams1987}
{Adams}, F.~C., {Lada}, C.~J., \& {Shu}, F.~H. 1987, \apj, 312, 788, \dodoi{10.1086/164924}

\bibitem[{{Adams} \& {Shu}(1985)}]{Adams1985}
{Adams}, F.~C., \& {Shu}, F.~H. 1985, \apj, 296, 655, \dodoi{10.1086/163483}

\bibitem[{{Aoyama} \& {Ikoma}(2019)}]{Aoyama2019}
{Aoyama}, Y., \& {Ikoma}, M. 2019, \apjl, 885, L29, \dodoi{10.3847/2041-8213/ab5062}

\bibitem[{{Ayliffe} \& {Bate}(2009)}]{Ayliffe2009}
{Ayliffe}, B.~A., \& {Bate}, M.~R. 2009, \mnras, 397, 657, \dodoi{10.1111/j.1365-2966.2009.15002.x}

\bibitem[{{Bae} {et~al.}(2022){Bae}, {Teague}, {Andrews}, {Benisty}, {Facchini}, {Galloway-Sprietsma}, {Loomis}, {Aikawa}, {Alarc{\'o}n}, {Bergin}, {Bergner}, {Booth}, {Cataldi}, {Cleeves}, {Czekala}, {Guzm{\'a}n}, {Huang}, {Ilee}, {Kurtovic}, {Law}, {Le Gal}, {Liu}, {Long}, {M{\'e}nard}, {{\"O}berg}, {P{\'e}rez}, {Qi}, {Schwarz}, {Sierra}, {Walsh}, {Wilner}, \& {Zhang}}]{Bae2022}
{Bae}, J., {Teague}, R., {Andrews}, S.~M., {et~al.} 2022, \apjl, 934, L20, \dodoi{10.3847/2041-8213/ac7fa3}

\bibitem[{{Benisty} {et~al.}(2021){Benisty}, {Bae}, {Facchini}, {Keppler}, {Teague}, {Isella}, {Kurtovic}, {P{\'e}rez}, {Sierra}, {Andrews}, {Carpenter}, {Czekala}, {Dominik}, {Henning}, {Menard}, {Pinilla}, \& {Zurlo}}]{Benisty2021}
{Benisty}, M., {Bae}, J., {Facchini}, S., {et~al.} 2021, \apjl, 916, L2, \dodoi{10.3847/2041-8213/ac0f83}

\bibitem[{Bezanson {et~al.}(2017)Bezanson, Edelman, Karpinski, \& Shah}]{Julia}
Bezanson, J., Edelman, A., Karpinski, S., \& Shah, V.~B. 2017, SIAM review, 59, 65.
\newblock \url{https://doi.org/10.1137/141000671}

\bibitem[{{Blandford} \& {Payne}(1982)}]{Blanford1982}
{Blandford}, R.~D., \& {Payne}, D.~G. 1982, \mnras, 199, 883, \dodoi{10.1093/mnras/199.4.883}

\bibitem[{{Cassen} \& {Moosman}(1981)}]{Cassen1981}
{Cassen}, P., \& {Moosman}, A. 1981, \icarus, 48, 353, \dodoi{10.1016/0019-1035(81)90051-8}

\bibitem[{{Chabrier} \& {Baraffe}(2000)}]{Chabrier2000}
{Chabrier}, G., \& {Baraffe}, I. 2000, \araa, 38, 337, \dodoi{10.1146/annurev.astro.38.1.337}

\bibitem[{{Chevalier}(1983)}]{Chevalier1983}
{Chevalier}, R.~A. 1983, \apj, 268, 753, \dodoi{10.1086/160997}

\bibitem[{{Choksi} \& {Chiang}(2024)}]{Choksi2024}
{Choksi}, N., \& {Chiang}, E. 2024, arXiv e-prints, arXiv:2403.10057, \dodoi{10.48550/arXiv.2403.10057}

\bibitem[{{Christiaens} {et~al.}(2024){Christiaens}, {Samland}, {Henning}, {Portilla-Revelo}, {Perotti}, {Matthews}, {Absil}, {Decin}, {Kamp}, {Boccaletti}, {Tabone}, {Marleau}, {van Dishoeck}, {G{\"u}del}, {Lagage}, {Barrado}, {Garatti}, {Glauser}, {Olofsson}, {Ray}, {Scheithauer}, {Vandenbussche}, {Waters}, {Arabhavi}, {Grant}, {Jang}, {Kanwar}, {Schreiber}, {Schwarz}, {Temmink}, \& {{\"O}stlin}}]{Christiaens2024}
{Christiaens}, V., {Samland}, M., {Henning}, T., {et~al.} 2024, arXiv e-prints, arXiv:2403.04855.
\newblock \doarXiv{2403.04855}

\bibitem[{{Cugno} {et~al.}(2024){Cugno}, {Patapis}, {Banzatti}, {Meyer}, {Dannert}, {Stolker}, {MacDonald}, \& {Pontoppidan}}]{Cugno2024}
{Cugno}, G., {Patapis}, P., {Banzatti}, A., {et~al.} 2024, arXiv e-prints, arXiv:2404.07086, \dodoi{10.48550/arXiv.2404.07086}

\bibitem[{Danisch \& Krumbiegel(2021)}]{Makie}
Danisch, S., \& Krumbiegel, J. 2021, Journal of Open Source Software, 6, 3349, \dodoi{10.21105/joss.03349}

\bibitem[{{Dominik} \& {Tielens}(1997)}]{Dominik1997}
{Dominik}, C., \& {Tielens}, A.~G.~G.~M. 1997, \apj, 480, 647, \dodoi{10.1086/303996}

\bibitem[{{Draine} \& {Lee}(1984)}]{Draine1984}
{Draine}, B.~T., \& {Lee}, H.~M. 1984, \apj, 285, 89, \dodoi{10.1086/162480}

\bibitem[{{Dullemond} \& {Dominik}(2005)}]{Dullemond2005}
{Dullemond}, C.~P., \& {Dominik}, C. 2005, \aap, 434, 971, \dodoi{10.1051/0004-6361:20042080}

\bibitem[{{Dullemond} {et~al.}(2012){Dullemond}, {Juhasz}, {Pohl}, {Sereshti}, {Shetty}, {Peters}, {Commercon}, \& {Flock}}]{Dullemond2012}
{Dullemond}, C.~P., {Juhasz}, A., {Pohl}, A., {et~al.} 2012, {RADMC-3D: A multi-purpose radiative transfer tool}, Astrophysics Source Code Library, record ascl:1202.015

\bibitem[{{Fung} {et~al.}(2019){Fung}, {Zhu}, \& {Chiang}}]{Fung2019}
{Fung}, J., {Zhu}, Z., \& {Chiang}, E. 2019, \apj, 887, 152, \dodoi{10.3847/1538-4357/ab53da}

\bibitem[{{Ghosh} \& {Lamb}(1978)}]{Ghosh1978}
{Ghosh}, P., \& {Lamb}, F.~K. 1978, \apjl, 223, L83, \dodoi{10.1086/182734}

\bibitem[{{Haffert} {et~al.}(2019){Haffert}, {Bohn}, {de Boer}, {Snellen}, {Brinchmann}, {Girard}, {Keller}, \& {Bacon}}]{Haffert2019}
{Haffert}, S.~Y., {Bohn}, A.~J., {de Boer}, J., {et~al.} 2019, Nature Astronomy, 3, 749, \dodoi{10.1038/s41550-019-0780-5}

\bibitem[{{Hartmann}(2009)}]{Hartmann2009}
{Hartmann}, L. 2009, {Accretion Processes in Star Formation: Second Edition} (Cambridge University Press)

\bibitem[{{Hashimoto} {et~al.}(2020){Hashimoto}, {Aoyama}, {Konishi}, {Uyama}, {Takasao}, {Ikoma}, \& {Tanigawa}}]{Hashimoto2020}
{Hashimoto}, J., {Aoyama}, Y., {Konishi}, M., {et~al.} 2020, \aj, 159, 222, \dodoi{10.3847/1538-3881/ab811e}

\bibitem[{{Hayashi}(1981)}]{Hayashi1981}
{Hayashi}, C. 1981, Progress of Theoretical Physics Supplement, 70, 35, \dodoi{10.1143/PTPS.70.35}

\bibitem[{{Helled} {et~al.}(2014){Helled}, {Bodenheimer}, {Podolak}, {Boley}, {Meru}, {Nayakshin}, {Fortney}, {Mayer}, {Alibert}, \& {Boss}}]{Helled2014}
{Helled}, R., {Bodenheimer}, P., {Podolak}, M., {et~al.} 2014, in Protostars and Planets VI, ed. H.~{Beuther}, R.~S. {Klessen}, C.~P. {Dullemond}, \& T.~{Henning}, 643--665, \dodoi{10.2458/azu_uapress_9780816531240-ch028}

\bibitem[{{Hern{\'a}ndez} {et~al.}(2007){Hern{\'a}ndez}, {Hartmann}, {Megeath}, {Gutermuth}, {Muzerolle}, {Calvet}, {Vivas}, {Brice{\~n}o}, {Allen}, {Stauffer}, {Young}, \& {Fazio}}]{Hernandez2007}
{Hern{\'a}ndez}, J., {Hartmann}, L., {Megeath}, T., {et~al.} 2007, \apj, 662, 1067, \dodoi{10.1086/513735}

\bibitem[{Hurwitz(1882)}]{Hurwitz1882}
Hurwitz, A. 1882, Zeitschrift f\"ur Math. u. Physik, 27, 86–101

\bibitem[{{Isella} {et~al.}(2019){Isella}, {Benisty}, {Teague}, {Bae}, {Keppler}, {Facchini}, \& {P{\'e}rez}}]{Isella2019}
{Isella}, A., {Benisty}, M., {Teague}, R., {et~al.} 2019, \apjl, 879, L25, \dodoi{10.3847/2041-8213/ab2a12}

\bibitem[{{Kanagawa} {et~al.}(2015){Kanagawa}, {Tanaka}, {Muto}, {Tanigawa}, \& {Takeuchi}}]{Kanagawa2015}
{Kanagawa}, K.~D., {Tanaka}, H., {Muto}, T., {Tanigawa}, T., \& {Takeuchi}, T. 2015, \mnras, 448, 994, \dodoi{10.1093/mnras/stv025}

\bibitem[{{Lada} \& {Wilking}(1984)}]{LadaWilk}
{Lada}, C.~J., \& {Wilking}, B.~A. 1984, \apj, 287, 610, \dodoi{10.1086/162719}

\bibitem[{{Lambrechts} \& {Lega}(2017)}]{Lambrechts2017}
{Lambrechts}, M., \& {Lega}, E. 2017, \aap, 606, A146, \dodoi{10.1051/0004-6361/201731014}

\bibitem[{{Lambrechts} {et~al.}(2019){Lambrechts}, {Lega}, {Nelson}, {Crida}, \& {Morbidelli}}]{Lambrechts2019}
{Lambrechts}, M., {Lega}, E., {Nelson}, R.~P., {Crida}, A., \& {Morbidelli}, A. 2019, \aap, 630, A82, \dodoi{10.1051/0004-6361/201834413}

\bibitem[{{Li} {et~al.}(2023){Li}, {Chen}, \& {Lin}}]{Li2023}
{Li}, Y.-P., {Chen}, Y.-X., \& {Lin}, D. N.~C. 2023, \mnras, 526, 5346, \dodoi{10.1093/mnras/stad3049}

\bibitem[{{Lovelace} {et~al.}(2011){Lovelace}, {Covey}, \& {Lloyd}}]{Lovelace2011}
{Lovelace}, R.~V.~E., {Covey}, K.~R., \& {Lloyd}, J.~P. 2011, \aj, 141, 51, \dodoi{10.1088/0004-6256/141/2/51}

\bibitem[{{Lunine} \& {Stevenson}(1982)}]{Lunine1982}
{Lunine}, J.~I., \& {Stevenson}, D.~J. 1982, \icarus, 52, 14, \dodoi{10.1016/0019-1035(82)90166-X}

\bibitem[{{Machida} {et~al.}(2008){Machida}, {Kokubo}, {Inutsuka}, \& {Matsumoto}}]{Machida2008}
{Machida}, M.~N., {Kokubo}, E., {Inutsuka}, S.-i., \& {Matsumoto}, T. 2008, \apj, 685, 1220, \dodoi{10.1086/590421}

\bibitem[{{Maeda} {et~al.}(2022){Maeda}, {Ohtsuki}, {Tanigawa}, {Machida}, \& {Suetsugu}}]{Maeda2022}
{Maeda}, N., {Ohtsuki}, K., {Tanigawa}, T., {Machida}, M.~N., \& {Suetsugu}, R. 2022, \apj, 935, 56, \dodoi{10.3847/1538-4357/ac7ddf}

\bibitem[{{Marley} {et~al.}(2007){Marley}, {Fortney}, {Hubickyj}, {Bodenheimer}, \& {Lissauer}}]{Marley2007}
{Marley}, M.~S., {Fortney}, J.~J., {Hubickyj}, O., {Bodenheimer}, P., \& {Lissauer}, J.~J. 2007, \apj, 655, 541, \dodoi{10.1086/509759}

\bibitem[{{Marley} {et~al.}(2021){Marley}, {Saumon}, {Visscher}, {Lupu}, {Freedman}, {Morley}, {Fortney}, {Seay}, {Smith}, {Teal}, \& {Wang}}]{Marley2021}
{Marley}, M.~S., {Saumon}, D., {Visscher}, C., {et~al.} 2021, \apj, 920, 85, \dodoi{10.3847/1538-4357/ac141d}

\bibitem[{{Mendoza} {et~al.}(2009){Mendoza}, {Tejeda}, \& {Nagel}}]{Mendoza2009}
{Mendoza}, S., {Tejeda}, E., \& {Nagel}, E. 2009, \mnras, 393, 579, \dodoi{10.1111/j.1365-2966.2008.14210.x}

\bibitem[{{Mosqueira} \& {Estrada}(2003{\natexlab{a}})}]{Mosqueira2003a}
{Mosqueira}, I., \& {Estrada}, P.~R. 2003{\natexlab{a}}, \icarus, 163, 198, \dodoi{10.1016/S0019-1035(03)00076-9}

\bibitem[{{Mosqueira} \& {Estrada}(2003{\natexlab{b}})}]{Mosqueira2003b}
---. 2003{\natexlab{b}}, \icarus, 163, 232, \dodoi{10.1016/S0019-1035(03)00077-0}

\bibitem[{{Paardekooper} {et~al.}(2023){Paardekooper}, {Dong}, {Duffell}, {Fung}, {Masset}, {Ogilvie}, \& {Tanaka}}]{Paardekooper2023}
{Paardekooper}, S., {Dong}, R., {Duffell}, P., {et~al.} 2023, in Astronomical Society of the Pacific Conference Series, Vol. 534, Protostars and Planets VII, ed. S.~{Inutsuka}, Y.~{Aikawa}, T.~{Muto}, K.~{Tomida}, \& M.~{Tamura}, 685, \dodoi{10.48550/arXiv.2203.09595}

\bibitem[{{Pollack} {et~al.}(1996){Pollack}, {Hubickyj}, {Bodenheimer}, {Lissauer}, {Podolak}, \& {Greenzweig}}]{Pollack1996}
{Pollack}, J.~B., {Hubickyj}, O., {Bodenheimer}, P., {et~al.} 1996, \icarus, 124, 62, \dodoi{10.1006/icar.1996.0190}

\bibitem[{{Quillen} \& {Trilling}(1998)}]{Quillen1998}
{Quillen}, A.~C., \& {Trilling}, D.~E. 1998, \apj, 508, 707, \dodoi{10.1086/306421}

\bibitem[{{Schulik} {et~al.}(2019){Schulik}, {Johansen}, {Bitsch}, \& {Lega}}]{Schulik2019}
{Schulik}, M., {Johansen}, A., {Bitsch}, B., \& {Lega}, E. 2019, \aap, 632, A118, \dodoi{10.1051/0004-6361/201935473}

\bibitem[{{Schulik} {et~al.}(2020){Schulik}, {Johansen}, {Bitsch}, {Lega}, \& {Lambrechts}}]{Schulik2020}
{Schulik}, M., {Johansen}, A., {Bitsch}, B., {Lega}, E., \& {Lambrechts}, M. 2020, \aap, 642, A187, \dodoi{10.1051/0004-6361/202037556}

\bibitem[{{Semenov} {et~al.}(2003){Semenov}, {Henning}, {Helling}, {Ilgner}, \& {Sedlmayr}}]{Semenov2003}
{Semenov}, D., {Henning}, T., {Helling}, C., {Ilgner}, M., \& {Sedlmayr}, E. 2003, \aap, 410, 611, \dodoi{10.1051/0004-6361:20031279}

\bibitem[{{Shakura} \& {Sunyaev}(1973)}]{Shakura1973}
{Shakura}, N.~I., \& {Sunyaev}, R.~A. 1973, \aap, 24, 337

\bibitem[{{Sun} {et~al.}(2024){Sun}, {Huang}, {Dong}, \& {Liu}}]{Sun2024}
{Sun}, X., {Huang}, P., {Dong}, R., \& {Liu}, S.-F. 2024, arXiv e-prints, arXiv:2406.09501.
\newblock \doarXiv{2406.09501}

\bibitem[{{Szul{\'a}gyi} {et~al.}(2016){Szul{\'a}gyi}, {Masset}, {Lega}, {Crida}, {Morbidelli}, \& {Guillot}}]{Szulagyi2016}
{Szul{\'a}gyi}, J., {Masset}, F., {Lega}, E., {et~al.} 2016, \mnras, 460, 2853, \dodoi{10.1093/mnras/stw1160}

\bibitem[{{Szul{\'a}gyi} \& {Mordasini}(2017)}]{Szulagyi2017}
{Szul{\'a}gyi}, J., \& {Mordasini}, C. 2017, \mnras, 465, L64, \dodoi{10.1093/mnrasl/slw212}

\bibitem[{{Takata} \& {Stevenson}(1996)}]{Takata1996}
{Takata}, T., \& {Stevenson}, D.~J. 1996, \icarus, 123, 404, \dodoi{10.1006/icar.1996.0167}

\bibitem[{{Tanigawa} {et~al.}(2012){Tanigawa}, {Ohtsuki}, \& {Machida}}]{Tanigawa2012}
{Tanigawa}, T., {Ohtsuki}, K., \& {Machida}, M.~N. 2012, \apj, 747, 47, \dodoi{10.1088/0004-637X/747/1/47}

\bibitem[{{Tanigawa} \& {Watanabe}(2002)}]{Tanigawa2002}
{Tanigawa}, T., \& {Watanabe}, S.-i. 2002, \apj, 580, 506, \dodoi{10.1086/343069}

\bibitem[{Taylor \& Adams(2024)}]{Taylor2024}
Taylor, A.~G., \& Adams, F.~C. 2024, Icarus, 116044, \dodoi{https://doi.org/10.1016/j.icarus.2024.116044}

\bibitem[{{Thanathibodee} {et~al.}(2019){Thanathibodee}, {Calvet}, {Bae}, {Muzerolle}, \& {Hern{\'a}ndez}}]{Thanathibodee2019}
{Thanathibodee}, T., {Calvet}, N., {Bae}, J., {Muzerolle}, J., \& {Hern{\'a}ndez}, R.~F. 2019, \apj, 885, 94, \dodoi{10.3847/1538-4357/ab44c1}

\bibitem[{{Turner} {et~al.}(2014){Turner}, {Lee}, \& {Sano}}]{Turner2014}
{Turner}, N.~J., {Lee}, M.~H., \& {Sano}, T. 2014, \apj, 783, 14, \dodoi{10.1088/0004-637X/783/1/14}

\bibitem[{{Ulrich}(1976)}]{Ulrich1976}
{Ulrich}, R.~K. 1976, \apj, 210, 377, \dodoi{10.1086/154840}

\bibitem[{{Weidenschilling}(1984)}]{Weidenschilling1984}
{Weidenschilling}, S.~J. 1984, \icarus, 60, 553, \dodoi{10.1016/0019-1035(84)90164-7}

\bibitem[{{Wolfram Research{,} Inc.}(2022)}]{Mathematica}
{Wolfram Research{,} Inc.} 2022, Mathematica, {V}ersion 13.2.
\newblock \url{https://www.wolfram.com/mathematica}

\bibitem[{{Zhu}(2015)}]{Zhu2015}
{Zhu}, Z. 2015, \apj, 799, 16, \dodoi{10.1088/0004-637X/799/1/16}

\bibitem[{{Zhu} {et~al.}(2018){Zhu}, {Andrews}, \& {Isella}}]{Zhu2018}
{Zhu}, Z., {Andrews}, S.~M., \& {Isella}, A. 2018, \mnras, 479, 1850, \dodoi{10.1093/mnras/sty1503}

\bibitem[{{Zhu} {et~al.}(2016){Zhu}, {Ju}, \& {Stone}}]{Zhu2016}
{Zhu}, Z., {Ju}, W., \& {Stone}, J.~M. 2016, \apj, 832, 193, \dodoi{10.3847/0004-637X/832/2/193}

\end{thebibliography}

\appendix

\section{Planck Weighted Quantities}\label{sec:bolokappa}

This Appendix determines the Planck mean opacity and an analogous Planck weighted mean attenuation factor. These results apply for a power-law form for the frequency dependent opacity, which can be written in the form 
$\kappa_\nu=\kappa_0(\nu/\nu_0)^\eta$. The index is expected to lie in the range $0\le\eta\le2$, where the value depends on both the grain composition and the wavelength range over which the power-law approximation is used (e.g., \citealt{Draine1984}, \citealt{Semenov2003}, and many others). 

\subsection{Planck Mean Opacity}\label{subsec:kPderiv}

The Planck mean opacity $\kappa_P$ is defined as 
\begin{equation}
    \kappa_P\equiv\int\displaylimits_0^\infty\!\! \kappa_\nu B_\nu(T)\,\de\nu\bigg/\!\!\int\displaylimits_0^\infty\!\! B_\nu(T)\,\de\nu\,,\label{eq:kappaPdef}
\end{equation}
where $B_\nu(T)$ is the usual blackbody function. The integrals can be evaluated using standard techniques \citep{AbramStegun1972} so that the Planck mean opacity has the form 
\begin{equation}
    \kappa_P=\kappa_0\Big(\frac{k_B T}{h\nu_0}\Big)^\eta\,\frac{\Gamma(4+\eta)\zeta_R(4+\eta)}{6\zeta_R(4)}\,,\label{eq:kappaP}
\end{equation}
where $\Gamma(z)$ is the gamma function and $\zeta_R(z)$ is the Riemann zeta function. {It is useful to separate the temperature dependence  and write $\kappa_P=b_\kappa T^\eta$, where $b_\kappa$ can be derived from Eq. \eqref{eq:kappaP}. }

\subsection{Mean Attenuation Coefficient}\label{subsec:alphaderiv}

It is useful to define a Planck-weighted mean attenuation coefficient $\alpha_P$ according to 
\begin{equation}
    \alpha_P\equiv\int\displaylimits_0^\infty\!\! B_\nu(T)\exp(-\kappa_\nu N_{\rm col})\,\de\nu\bigg/\!\!\int\displaylimits_0^\infty\!\! B_\nu(T)\,\de\nu\,.\label{eq:alphadef}
\end{equation}
In this instance, we specialize to the case where the index $\eta=1$ such that the opacity is a linear function of frequency.  

The denominator of Eq.~\eqref{eq:alphadef} has already been computed above and elsewhere. We define  $a\equiv\kappa_0N_{\rm col}/\nu_0$ and $b\equiv h/(k_B T)$, so that the numerator becomes
\begin{equation}
   \frac{2h}{c^2} \int\displaylimits_0^\infty\!\! \nu^3 \frac{\exp(-(a+b)\nu)}{1-\exp(-b\nu)}\,\de\nu\,.
\end{equation}
Replacing the integration variable with $x=b\nu$, we can then rewrite this integral in the form  
\begin{equation}
     \frac{2h}{c^2b^4}\int\displaylimits_0^\infty\!\! x^3\exp\left[-\frac{a+b}{b}x\right][1-\exp(-x)]^{-1}\,\de x\,. \label{eq:alphanumform}
\end{equation}
The resulting integral is now in a form equivalent to the integral definition of the Hurwitz zeta function \citep{Hurwitz1882}, i.e., 
\begin{equation}
    \zeta_H(s, a)=\frac{1}{\Gamma(s)}\!\!\int\displaylimits_0^\infty\!\!\frac{x^{s-1}e^{-ax}}{1-e^{-x}}\de x\,.
\end{equation}
The expression for the numerator becomes 
\begin{equation}
   \frac{2h}{c^2b^4}\Gamma(4)\,\zeta_H\!\!\left(4, 1+\frac{a}{b}\right)\,.
\end{equation}
Replacing $a$ and $b$ with their definitions, the attenuation coefficient can be written as
\begin{equation}\label{eq:alpha}
    \alpha_P=\zeta_H\Big(4, 1+\frac{\kappa_0 N_{\rm col} k_B T}{h\nu_0}\Big)\Big/\zeta_R(4)\,.
\end{equation}

\end{document}